\documentstyle[11pt,amsmath,amssymb]{article}

\newcommand{\bbbone}{\mathchoice {\rm 1\mskip-4mu l} {\rm 1\mskip-4mu l}
{\rm 1\mskip-4.5mu l} {\rm 1\mskip-5mu l}}
\newcommand{\scalprod}[2]{\left\langle {#1}, {#2}\right\rangle}
\newcommand{\dom}{{\cal D}}

\renewcommand{\Im}{{\rm Im}}
\newcommand{\fer}[1]{(\ref{#1})}
\newcommand{\ran}{{\rm Ran\,}}

\newcommand{\h}{{\cal H}}
\newcommand{\cx}{{\mathbb C}}
\newcommand{\rx}{{\mathbb R}}
\newcommand{\r}{{\rm R}}
\newcommand{\s}{{\rm S}}
\newcommand{\lless}{<\!\!<}
\newcommand{\ggeq}{>\!\!>}

\newcommand{\Lbar}{{\overline L}}
\renewcommand{\d}{{\rm d}}

\newcommand{\pbar}{\,\overline{P}\,}
\newcommand{\mm}{{\frak M}}
\renewcommand{\aa}{{\frak A}}

\newcommand{\e}{{\rm e}}

\renewcommand{\tilde}[1]{\widetilde{#1}}
\renewcommand{\i}{{\rm i}}
\newcommand{\psiref}{\psi_{\rm ref}}

\renewcommand{\c}{{\cal C}}
\newcommand{\Deltamax}{\Delta_{\rm max}}

\newcommand{\unity}{\bbbone}
\newcommand{\ket}[1]{\vert{#1}\rangle}

\newcommand{\R}{\rx}

\newcounter{resultcounter}[section]

\newtheorem{theorem}[resultcounter]{Theorem}
\newtheorem{lemma}[resultcounter]{Lemma}
\newtheorem{proposition}[resultcounter]{Proposition}

\begin{document}

\title{A resonance theory for open quantum systems\\
 with time-dependent dynamics\\
 \ \ \\
\large\it  Dedicated to J\"urg Fr\"ohlich and Tom Spencer\\
with our respect and affection}

\author{
Marco Merkli\footnote{Department of Mathematics and Statistics, Memorial University of Newfoundland, St. John's, NL, Canada, A1C 5S7, merkli@math.mun.ca, http://www.math.mun.ca/\~{}merkli/\ . Suppported by the Natural Sciences and Engineering Research Council of Canada (NSERC) under Grant 205247.} \quad and \quad Shannon Starr\footnote{Department of Mathematics, University of Rochester, Rochester, NY 14627, USA, sstarr@math.rochester.edu, http://www.math.rochester.edu/people/faculty/sstarr/ .  Supported in part by a U.S. National Science Foundation grant, DMS-0706927.}
}
\date{\today}

\maketitle

\begin{abstract}
We develop a resonance theory to describe the evolution of open systems with time-dependent dynamics. Our approach is based on piecewise constant Hamiltonians: we represent the evolution on each constant bit using a recently developed dynamical resonance theory, and we piece them together to obtain the total evolution. The initial state corresponding to one time-interval with constant Hamiltonian is the final state of the system corresponding to the interval before. This results in a non-markovian dynamics. 
We find a representation of the dynamics in terms of resonance energies and resonance states associated to the Hamiltonians, valid for all times $t\geq 0$ and for small (but fixed) interaction strengths. The representation has the form of a path integral over resonances.
We present applications to a spin-fermion system, where the energy levels of the spin may undergo rather arbitrary crossings in the course of time. In particular, we find the probability for transition between ground- and excited state at all times. 
\end{abstract}

\setcounter{page}{1}
\setcounter{section}{1}
\setcounter{section}{0}

\section{Introduction and outline of main results}
\label{introsect}

We study the evolution of an open quantum system $\s$ in contact with a quantum heat reservoir $\r$. The Hamiltonian of $\s$, as well as the interaction between the two systems is time-dependent. Our goal is to derive the form of the reduced dynamics of $\s$ for all times $t\geq 0$ and for small (but fixed) values of the coupling constant governing the strength of the interaction. An analysis of this kind  for time-independent dynamics has been carried out in \cite{MSB1,MSB2}. The approach adopted in the present work is based on the methods developed in these references, which in turn are extensions of a recent theory of quantum resonances for the analysis of large-time asymptotics of open quantum systems \cite{JP2,BFSrte,FM,JP3,MMS1,MMS2} (see also the references in these works for further literature). 

Within the context of the present paper, the long-time asymptotics has been examined for time-dependent dynamics in the following settings: in \cite{FMSU} for interaction operators having some limit as $t\rightarrow\infty$, in \cite{FMSU,AF} for periodic interactions (using algebraic scattering theory and Floquet theory, respectively), in \cite{BJM1,BJM2} for piecewise constant dynamics and markovian reservoirs (repeated interaction systems) and in \cite{A,AF1} for adiabatic dynamics. All these works are concerned with the approach of the system to an asymptotic state and with the thermodynamic properties of the latter. In contrast, in the present paper, we examine the dynamics of the open system for all times and for rather arbitrary time-dependences of the dynamics (not necessarily leading to an asymptotic state of the system).

We develop a resonance theory for Hamiltonians of the form
\begin{equation}
H(t)=H_\s(t)+H_\r+\lambda(t) v(t),
\label{i1}
\end{equation}
where $H_\s(t)$ and $H_\r$ are the Hamiltonians of $\s$ and $\r$ respectively, $\lambda(t)$ is a coupling constant, and $v(t)$ is an interaction operator. We base our approach on piecewise constant Hamiltonians of the form \fer{i1}, meaning that the Heisenberg dynamics of an observable $A$ is given by
\begin{equation}
\alpha_N(A) = \e^{\i t_1 H^1}\cdots \e^{\i t_N H^N} A \e^{-\i t_N H^N} \cdots \e^{-\i t_1 H^1},
\label{i2}
\end{equation}
where $t_j>0$ and 
\begin{equation}
H^j = H_\s^j+H_\r +\lambda_j v^j.
\label{i3}
\end{equation}
The dynamics \fer{i2} describes sudden changes in parameters of $\s$ and the interaction, and it has its own interest. A piecewise constant dynamics may also be viewed as an approximation of a continuous dynamics, in the appropriate limit $t_j\rightarrow 0$ and $N\rightarrow\infty$. We illustrate both these settings on concrete models in Section \ref{sectappl}.

\medskip
The space of pure states $\h_\s$ of the open system $\s$ is a finite-dimensional Hilbert space, and its Hamiltonian $H_\s^j$ is an arbitrary self-adjoint operator on $\h_\s$. We model the reservoir by a spatially infinitely extended ($\rx^3$) gas of free Fermions in equilibrium at temperature $T>0$. The Hamiltonian of $\r$ is given by 
\begin{equation}
H_\r = \int_{\rx^3} |k|^2 a^*(k)a(k) \d^3k,
\label{i4}
\end{equation}
where the $a(k)$ and $a^*(k)$ are fermionic annihilation and creation operators, satisfying the standard canonical anti-commutation relations $\{a(k),a^*(l)\}=\delta(k-l)$, see e.g. \cite{BR}. We understand that in \fer{i4}, and for all other quantities involving $\r$, the thermodynamic (infinite volume, continuous mode) limit has to be taken (this is the so called Araki-Wyss representation \cite{AW}, see also Section \ref{awapp} for further details). The equilibrium state of $R$ is the quasi-free state determined by the two-point function
\begin{equation}
\omega_{\r,\beta}\big(a^*(k)a(l)\big) = \frac{\delta(k-l)}{\e^{\beta |k|^2}+1},
\label{i4.1}
\end{equation}
where $\beta=1/T$, see e.g. \cite{BR}.

The interaction operator is a sum of terms of the form
\begin{equation}
v^j= G^j\otimes \phi(g^j),
\label{i5}
\end{equation}
where $G^j$ is any self-adjoint operator on $\h_\s$, and 
\begin{equation}
\phi(g^j) = \frac{1}{\sqrt 2}[a^*(g^j) +a(g^j)]
\label{i6}
\end{equation}
is the field operator smoothed out with a function $g^j\in L^2(\rx^3,\d^3 k)$, called a {\it form factor}. Here, the smoothed-out creation and annihilation operators are defined by
\begin{equation}
a^*(g) =\int_{\rx^3} g(k) a^*(k)\d^3 k,\qquad a(g) =\int_{\rx^3} \overline g(k) a(k)\d^3 k
\label{i7}
\end{equation}
(we take annihilation operators to be anti-linear in their arguments). Interactions of the form \fer{i5} induce processes of absorption and emission of quanta of $\r$ by the system $\s$.

Our approach uses a spectral deformations (generated by translation in the energy variable in a suitable Hilbert space). This method necessitates certain regularity of the form factors $g^j$. We represent $g^j(r,\sigma)$ in spherical coordinates $(r,\sigma)\in\rx_+\times S^2$ and denote by $\overline{g^j}$ its complex conjugate. 
\begin{itemize}
\item[{\bf (R)}]{\it Assumption on regularity (translation analyticity) of form-factors.}
The maps
\begin{equation}
\rx\times S^2\ni (u,\sigma)\mapsto \alpha(u) \sqrt{\frac{|u|^{1/4}}{\e^{-\beta u}+1}}\left\{
\begin{array}{ll}
g^j(\sqrt u,\sigma), & \mbox{if $u\geq 0$},\\
\overline{g^j}(\sqrt{-u},\sigma), & \mbox{if $u<0$},
\end{array}
\right.
\label{ancond}
\end{equation}
where $\alpha(u)=1, \e^{-\beta u}$, extend analytically (in $u$) to maps from $(u,\sigma)\in \{ z\in\cx\ :\ |\Im z|<\delta\}\times S^2$ to $L^2(\rx\times S^2,\d u\d\sigma)$, for some $\delta>0$.
\end{itemize}
An example of a form factor satisfying this assumption is $g(k) = |k|^{-1/2}\e^{-|k|^2}$.

The next assumption concerns the ``complete splitting of resonances''. We make it merely for the purpose of a lighter exposition of our results.  
Fix $j$ and let $e\in\{ E-E'\ :\ E,E'\in {\rm spec}(H_\s^j)\}$ be an energy difference of the system $\s$. In the resonance approach, the evolution of $\s$ is described by resonance energies $\varepsilon$, which are complex in general and reflect the non-unitary (irreversible) character of the reduced dynamics of $\s$. As the interaction between $\s$ and $\r$ is turned on, resonance energies $\varepsilon$ bifurcate out of each (real) energy difference $e$. The total multiplicity of resonance energies bifurcating out of a given $e$ equals ${\rm mult}(e)$ (the multiplicity of $e$ viewed as an eigenvalue of the operator $H_\s^j\otimes\bbbone_\s - \bbbone_\s\otimes H_\s^j$ acting on $\h_\s\otimes\h_\s$). We label the resonances by $r=(e,s)$, where $e$ denotes the origin of bifurcation and $1\leq s\leq {\rm mult}(e)$ counts {\it distinct} resonance energies. 
\begin{itemize}
\item[{\bf (S)}]
{\it Assumption on complete splitting of resonances.} At all time-steps $j$, for $\lambda_j\neq 0$ there are ${\rm mult}(e)$ distinct resonance energies $\varepsilon$ associated to each eigenvalue difference $e$.
\end{itemize}

One can deal equally well with degenerate resonance energies by adapting the arguments of \cite{MSB2} to the time-dependent case. Assumption (S) can be verified by using a perturbative analysis of the resonance energies (see \fer{i8} below).

\subsection{Dynamics of $\s$}

Let us explain our main result on the dynamics of the system $\s$, whose precise statement is given in Theorem \ref{mmthm1} below. We consider initial states of the form $\omega_0 = \omega_{\s,0}\otimes\omega_{\r,\beta}$, where $\omega_{\s,0}$ is an arbitrary state of $\s$, and $\omega_{\r,\beta}$ is given by \fer{i4.1}.\footnote{Our theory works as well for states which are local perturbations of such states, but we restrict our exposition to product initial states.} Theorem \ref{mmthm1} gives the following representation of the evolution of the average of an observable $A$ of the system $\s$ in the initial state $\omega_0$.
\begin{equation}
\omega_0\big(\alpha_N(A)\big) = \sum_{r_1,\ldots,r_N} \e^{\i\sum_{j=1}^N t_j\varepsilon^j(r_j)} \rho_{r_1,\ldots,r_N}(A) +O(\max_j|\lambda_j|\big).
\label{ii1}
\end{equation}
The sum is taken over indices $r_j=(e,s)$ which label the resonance energies $\varepsilon^j(r_j)$ associated to the system at step $j$. The $\rho_{r_1,\ldots,r_N}$ are linear functionals on the algebra of observables $\mm_\s={\cal B}(\h_\s)$ (bounded operators). Both $\varepsilon^j(r_j)$ and $\rho_{r_1,\ldots,r_N}$ depend on $\lambda_j$, and the remainder term depends on $N$ as well, but is uniform in the $t_j>0$ (it also depends on the interaction $v^j$, \fer{i5}). The resonance energies have the expansion 
\begin{equation}
\varepsilon^j = e +\lambda_j^2\delta^j +O(\lambda_j^4)
\label{i8}
\end{equation} 
for small $\lambda_j$. Here, the $\delta^j$ are eigenvalues of an operator $\Lambda^j(e)$ acting on the doubled space $\h_\s\otimes\h_\s$, called the {\it level shift operator} associated to $e$ at time-step $j$ (see the definition \fer{mm10}). We have $\Im\delta^j\geq 0$.\footnote{This can be seen by direct calculation in concrete models, and it can also be derived from general considerations, see as well \cite{MMS2}.} 

The functionals $\rho_{r_1,\ldots,r_N}$ can be expressed as
\begin{equation}
\rho_{r_1,\ldots,r_N}(A)={\tt P}\big(\Pi_0(r_1,\ldots,r_N )\  A\otimes\bbbone_\s\big),
\label{i9}
\end{equation}
where ${\tt P}$ is a linear functional on the algebra $\mm_\s\otimes\bbbone_\s$ acting on the doubled space $\h_\s\otimes\h_\s$. ${\tt P}$ depends on the initial state $\omega_0$ only. $\Pi_0$ is a product of ``transition amplitudes'' associated to $r_1,\ldots,r_N$,
\begin{equation}
\Pi_0(r_1,\ldots,r_N) = \left[\prod_{j=1}^{N-1} \scalprod{\widetilde\eta^j(r_j)}{\eta^{j+1}(r_{j+1})}\right] |\eta^1(r_1)\rangle\langle\widetilde\eta^N(r_N)|.
\label{i10}
\end{equation}
Here, the $\eta^j(r)$, $\widetilde\eta^j(r)\in\h_\s\otimes\h_\s$ are (the lowest order contributions of) {\it resonance eigenvectors}, defined by 
\begin{equation}
\Lambda^j(e)\eta^j(r) = \varepsilon^j(r)\eta^j(r) \quad \mbox{and} \quad [\Lambda^j(e)]^*\widetilde\eta^j(r) = \overline{\varepsilon^j(r)}\widetilde\eta^j(r)
\label{i11}
\end{equation}
and normalized as 
\begin{equation}
\scalprod{\eta^j(r)}{\widetilde\eta^j(r)}=1,\qquad \scalprod{\eta^j(r)}{\widetilde\eta^j(r')}=0 \mbox{\ if $r\neq r'$}
\label{i12}
\end{equation}
(recall that $r=(e,s)$). In \fer{i11}, $[\Lambda^j(e)]^*$ denotes the adjoint operator of $\Lambda^j(e)$.

\medskip
{\it Discussion of \fer{ii1}.\ }  -- At each moment when the Hamiltonian changes, the system starts a new dynamical process with an initial condition determined by the final state of the previous process. Note that even if we start in an unentangled (product) state of $\s+\r$, already after the first bit of interaction the state will become entangled. Since $\s$ interacts with the same reservoir at each time-step, the dynamics is not markovian. The cumulative effect of the interactions is encoded in the functionals $\rho_{r_1,\ldots,r_N}$ and the product of increments of the dynamics $\e^{\i t_1 \varepsilon^1(r_1)}\cdots \e^{\i t_N \varepsilon^N(r_N)}$.

-- The case of a time-independent dynamics can be recovered from \fer{ii1} as follows. If $H^j$ is close to $H^{j+1}$, then $\eta^j(r)$ is close to $\eta^{j+1}(r)$ and the transition amplitude 
\begin{equation}
\scalprod{\widetilde\eta^j(r_j)}{\eta^{j+1}(r_{j+1})} \approx \scalprod{\widetilde\eta^j(r_j)}{\eta^j(r_{j+1})}
\label{i21}
\end{equation}
is very small unless $r_j=r_{j+1}$, in which case it is roughly unity (see \fer{i12}). 
We can view the sequence $r_1,\ldots,r_N$ as a ``path'' of resonances: the system hops from resonance $r_j$ to resonance $r_{j+1}$ as time passes the moment $t_1+\ldots+t_j$. Thus for small differences $H^j-H^{j+1}$, the main contribution to the sum in \fer{ii1} comes from the constant paths $r_j=r=\rm const$, with associated propagator $\e^{\i t\varepsilon(r)}$. In this limit of a time-independent Hamiltonian, \fer{ii1} reduces to the dynamics of $\s$ derived in \cite{MSB1}.

-- If the interaction $v^j$ is energy exchanging then it typically drives the total system $\s+\r$ to its equilibrium state at a relaxation rate $1/\tau^j_{\rm therm}$. In the regime $t_j\gtrsim \tau^j_{\rm therm}$ one then expects to find the system after each bit of constant interaction in equilibrium relative to the dynamics at that moment. This is an adiabatic process during which the state of the system follows its instantaneous equilibrium state, see also \cite{A,AF1}.

\subsection{Applications}
\label{sectappl}

We consider a spin-$\frac 12$ particle subject to a time-dependent Hamiltonian, coupled to a thermal Fermi field. The space of pure states of the spin (system $\s$) is $\cx^2$, and the Hamiltonian at time-step $j$ is given by 
\begin{equation}
H_\s^j= \frac{\Delta^j}{2} \sigma_z = \frac{\Delta^j}{2}
\left[
\begin{array}{cc}
1 & 0\\
0 & -1
\end{array}
\right],
\label{mmr1}
\end{equation}
where $\Delta^j\in\rx$ is the energy level spacing. At times $j$ when $\Delta^j$ switches its sign we say we have a level crossing. The interaction of $\s$ with $\r$ is given by $\lambda v$ (constant in time), where $\lambda$ is a coupling constant, and (recall \fer{i6})
\begin{equation}
v = \sigma_x\otimes\phi(g) = 
\left[
\begin{array}{cc}
0 & 1\\
1 & 0
\end{array}
\right]  \otimes \frac{1}{\sqrt 2}[a^*(g) + a(g)].
\label{mmr2}
\end{equation}
The parameter regime $\lambda^2\lless \min_j|\Delta^j|$ describes well separated resonances, while if $\lambda^2$ is of the order of the $\Delta^j$, or if $\lambda^2\ggeq \min_j|\Delta^j|$, then we have ``overlapping resonances''. 

\subsubsection{Regime of overlapping resonances}
Our goal is to analyze the dynamics of the system for small but independent values of $\lambda$ and $\Delta^j$. In Theorem \ref{thmmmr1} and Proposition \ref{proptranscoeff} we give the explicit form of the resonance data $\varepsilon^j(r), \eta^j(r), \widetilde\eta^j(r)$ for this model, as well as the transition amplitudes $\scalprod{\widetilde\eta^j(r)}{\eta^{j+1}(r')}$ (recall \fer{i10}). As an illustration of these results, we present here the case of a single sudden level crossing and the limit of continuous time-dependent dynamics.

\medskip
{\bf Single sudden level crossing.\ }
Consider a single sudden level crossing at time $t_{\rm c}$, whose evolution is generated by \fer{mmr2} and 
\begin{equation}
H^t_\s = \frac{\Delta_1}{2}\sigma_z,\ 0\leq t\leq t_{\rm c},\qquad H^t_\s = -\frac{\Delta_2}{2}\sigma_z,\ t_{\rm c}<t,
\label{mmr41}
\end{equation}
where $\Delta_1,\Delta_2>0$. Denote by $p_{\rm ge}(t)$ the probability that the system $\s$ is at time $t$ in the excited state of $H^t_\s$, while at time zero it started off in the ground state of $H^{t=0}_\s$. (By ground state we mean the state with lowest energy.) We show in Section \ref{subsubsslc} that for independently small values of $\lambda$ and $\Deltamax=\max\{\Delta_1,\Delta_2\}$, we have 
\begin{equation}
p_{\rm ge}(t) = \frac{1}{2}
\left\{
\begin{array}{ll}
1-\e^{\i t\varepsilon}  & \mbox{if $0\leq t<t_{\rm c}$}, \\
1+\e^{\i t\varepsilon}  & \mbox{if $t>t_{\rm c}$},\\
\end{array}
\right\}
+O(|\lambda|+\Delta_{\rm max}),
\label{mmr45.1}
\end{equation}
where 
\begin{equation}
\varepsilon = \i\pi\lambda^2\gamma_0 +O\big(\lambda^2(|\lambda|+\Deltamax)\big)
\label{i20}
\end{equation}
with 
\begin{equation}
\gamma_0= \lim_{r\rightarrow 0+} \frac{\sqrt{r}}{2}\int_{S^2} \d\sigma|g(\sqrt{r},\sigma)|^2.
\label{mmr5'}
\end{equation}
It is assumed here that $0<\gamma_0<\infty$, which amounts to an infra-red ($|k|\sim 0$) behaviour $g(|k|,\sigma)\sim |k|^{-1/2}$ in three space dimensions (spherical coordinates; see also Assumption (R) in Section \ref{introsect}). Formula \fer{mmr45.1} shows in particular that at $t_{\rm c}$ the probability jumps up by an amount
\begin{equation}
\delta= \e^{\i t_{\rm c}\varepsilon}+O(|\lambda|+\Delta_{\rm max}) = \e^{-\pi\gamma_0\lambda^2 t_{\rm c}[1+O(|\lambda|+\Delta_{\rm max})]} +O(|\lambda|+\Delta_{\rm max}).
\label{mmr45'}
\end{equation}
This is in part explained by the fact that the excited state itself jumps at $t_c$ from $\ket{+}$ to $\ket{-}$.
As $t\rightarrow\infty$, $p_{\rm ge}(t)$ approaches $1/2+O(|\lambda|+\Delta_{\rm max})$, which is the probability of finding $\s$ in the excited state when the total system $\s+\r$ is in equilibrium, provided $\lambda$ and $\Delta_{\rm max}$ are small. This is the correct value of this asymptotic probability, since the system exhibits return to equilibrium.

In a time-dependent setting where energy levels of a quantum system are brought close together (but do not cross) in the course of time, say due to some external forcing, a transition from one energy state to another is called a Landau-Zener transition. Landau-Zener theory is important in physics and chemistry, see e.g. \cite{HJ,SO}. The influence of dissipation on Landau-Zener transitions has been studied for different systems and transition probabilities similar to \fer{mmr45.1} have been calculated \cite{PR,WSKHK}. We mention that our method allows for crossing of the energy levels in the course of time.

\medskip
{\bf Continuum limit.\ } We investigate the continuum limit of the model \fer{mmr1} by setting $t_j=j \frac{t}{N}$, $j=1,\ldots,N$, with fixed $t$ and letting $N\rightarrow\infty$. Our goal is to find the limit of the dominant contribution to the dynamics, given by the sum in \fer{ii1}. We do not control the remainder term in \fer{ii1} uniformly in $N$ in this paper (work on this is in progress).

We take $\Delta^j = \Delta(j\frac tN)$, where $\Delta(t)$ is a continuously differentiable function of $t\geq 0$ (of course, one can deal with less regular $\Delta$ if desired), and we define $\tau^j$ and $\sigma$ by
\begin{equation}
\tau^j = \frac{\Delta^j}{\sigma}, \qquad \sigma= \frac{\pi}{2}\lambda^2\gamma_0,
\label{mmr30}
\end{equation}
where $\gamma_0$ is given in \fer{mmr5'} above. For simplicity of the exposition, we will assume in what follows that $0<\tau_{\rm max}<1$, 
where $\tau_{\rm max}:=\frac{\Delta_{\rm max}}{\sigma}$ and $\Delta_{\rm max} = \sup_{t\geq 0}|\Delta(t)|$. This regime is interesting since it accommodates the situation of level crossings ($\Delta(t)=0$) while $\lambda$ is fixed.

We have seen above (see \fer{i21}) that if $\Delta^{j+1}-\Delta^j$ is small, then the transition amplitude $\scalprod{\widetilde\eta^j(r)}{\eta^{j+1}(r')}$ associated to a jump ($r\neq r'$) in the resonance path at $j$ is small. We show in the proof of Theorem \ref{contlimthm} that this amplitude is at most of the size 
\begin{equation}
\tau'_{\rm max}:=\sup_{t\geq 0} |\tau'(t)|.
\label{mmr50.1}
\end{equation}
(The $'$ here means derivative.) The sum over all paths in \fer{ii1} can be written as a sum over all paths with $k$ jumps, $k=0,\ldots,N-1$. This sum becomes an infinite series in the continuous time limit, and the summand associated to a path with $k$ jumps is of the order of $(\tau'_{\rm max})^k$. For small $\tau_{\rm max}'$, one can thus (rigorously) approximate the series by the first few terms. In Theorem \ref{contlimthm} we show that the continuous time  limit of the sum in \fer{ii1} is given by
\begin{eqnarray}
\lefteqn{\sum_{r=1}^4 \e^{\i\int_0^t\varepsilon(s,r)\d s} w(0,t,r) \scalprod{\psi_0}{B\eta(0,r)}\scalprod{\widetilde\eta(t,r)}{A\psiref}}\label{27}\\
&+&\int_0^t \e^{\i\int_0^s\varepsilon(s',3)\d s' +\i\int_s^t \varepsilon(s',4)\d s'}\ w(0,s,3)\frac{y_+(s)y_-'(s)}{1+y_+(s)^2} w(s,t,4) \d s\nonumber\\
&&\quad \times \scalprod{\psi_0}{B\eta(0,3)}\scalprod{\widetilde\eta(t,4)}{A\psiref}\nonumber\\
&+&\int_0^t \e^{\i\int_0^s\varepsilon(s',4)\d s' +\i\int_s^t \varepsilon(s',3)\d s'}\ w(0,s,4)\frac{y_-(s)y_+'(s)}{1+y_-(s)^2} w(s,t,3) \d s\nonumber\\
&&\quad \times \scalprod{\psi_0}{B\eta(0,4)}\scalprod{\widetilde\eta(t,3)}{A\psiref}\nonumber\\
&+&O\big([\tau'_{\rm max} t]^2\, \e^{2C\tau'_{\rm max}t}\big).  \nonumber
\end{eqnarray}
Here, $\varepsilon(t,r)$ and $\eta(t,r)$, $\widetilde \eta(t,r)$ are the resonance energies and resonance vectors at time $t\geq 0$ (with $r=1,\ldots,4$; see Section \ref{subsubcontlimit} for their explicit form). The functions $w$ and $y_\pm$ are associated to the transition amplitudes, $\psi_0$ is the (Gelfand-Naimark-Segal) vector representative of the initial state of $\s$ (represented in the Hilbert space $\h_\s\otimes\h_\s$), $\psiref$ is the vector representing the trace state of $\s$, and $B\in\bbbone_\s\otimes \mm_\s$ is the unique operator satisfying $\psi_0=B\psiref$. We refer the reader to Section \ref{subsubcontlimit} for the explicit expressions of all quantities involved in \fer{27}.

The sum in \fer{27} is the contribution coming from the constant paths, while the two integrals come from paths having a single jump (taking place at the integration variable $s$). Both these integrals are of the order of $\tau'_{\rm max}$. Naturally, the Riemann sum in the propagator of \fer{ii1} has now become an integral, and the product of the transition coefficients has turned into the factors $w$ (associated to products without jumps) and the fractions involving the $y_\pm$ (associated with single jumps). The remainder term contains the contributions of all paths with more than one jump. In fact, it is not very hard to find the continuous time limit of all paths with arbitrarily many jumps. We do not present the corresponding formulas in this work since they are rather cumbersome.

\subsubsection{Regime of isolated resonances}

The regime of isolated resonances can be obtained as a limit ($\tau^j\ggeq 1$) of the case of overlapping resonances,
but not the other way around.
More specifically, the expressions for the resonance energies and resonance vectors obtained for overlapping resonances (see Theorem \ref{thmmmr1}) are still valid once 
equations \fer{mmr5'} and \fer{mmr30} are replaced by
\begin{align}
\gamma(\Delta^j) 
&= \frac{\sqrt{|\Delta^j|}}{2} \int_{S^2}  \d \sigma  \left|g\left(\sqrt{|\Delta^j|},\sigma\right)\right|^2,\label{ssr1}\\
\sigma(\Delta^j)
&= \frac{\pi}{2} \lambda^2 \gamma(\Delta^j).
\label{ssr2}
\end{align}
We present the explicit form of the resonance data in Theorem \ref{thmssr1}.

\medskip
{\bf Single sudden level crossing.\ }
With the same set-up as before, but now assuming that $\Delta^1,\Delta^2>0$ are fixed independently of $\lambda$,
\begin{equation}
p_{\rm ge}(t) = \frac{1 - e^{-\pi^2 \lambda t \gamma(\Delta^1)}}
{e^{\beta \Delta^1} + 1} + O(|\lambda|)
\label{ssr4}
\end{equation}
for $t<t_c$ and 
\begin{equation}
p_{\rm ge}(t) = \frac{1}{e^{-\beta \Delta^2} + 1} + \frac{e^{-\pi \lambda^2 t_c \gamma(\Delta^1)}}{e^{-\beta \Delta^1}+1}
\left[e^{-\pi \lambda^2 (t-t_c) \gamma(\Delta^2)} - \frac{1-e^{-\beta(\Delta^1+\Delta^2)}}{e^{-\beta \Delta^2} + 1}\right]
+ O(|\lambda|)
\label{ssr5}
\end{equation}
for $t>t_c$.
Therefore, there is a jump up at $t_c$ equal to $e^{-\pi^2 \lambda t_c \gamma(\Delta^1)} + O(|\lambda|)$,
in perfect analogy to the situation before.

\bigskip
\noindent
{\bf Acknowledgements.\ } Parts of this work have been carried out during visits of each author to the other's institution. We are grateful for the support of the Mathematics Departments of Memorial University and of the University of Rochester, and to the NSERC and the NSF for their support. We also thank the referees for pointing out some typographical errors.

\section{Dynamical resonance theory}

We give a precise definition of the model in Section \ref{ssdofmodel}. In Section \ref{ssii1} we prove formula \fer{ii1}. The main result is Theorem \ref{mmthm1}.

\subsection{Description of the model}
\label{ssdofmodel}

The Hilbert space of states of the system $\s+\r$ is given by
\begin{equation}
\h =\h_\s\otimes\h_\s\otimes{\cal F},
\label{a1}
\end{equation}
where $\h_\s=\cx^d$ is the Hilbert space of pure states of $\s$, and 
\begin{equation}
{\cal F}={\cal F}\big(L^2({\mathbb R}\times S^2, \d u\times\d\Sigma)\big)
\label{10}
\end{equation}
is the fermionic Fock space over the one-particle space $L^2({\mathbb R}^3\times S^2, \d u\times\d\Sigma)$. $\h$ is the (Gelfand-Naimark-Segal) representation Hilbert space associated to the reference state 
\begin{equation}
\omega_{\rm ref} = \omega_{\s,\rm ref}\otimes\omega_{\r,\beta}.
\label{a2}
\end{equation}
Here, $\omega_{\s,\rm ref}$ is the trace state on the $C^*$ algebra of observables $\aa_\s= {\cal B}(\h_\s)$ (bounded operators on $\h_\s$), and $\omega_{\r,\beta}$ is the equilibrium state of the infinitely extended free Fermi gas (see also \fer{i4.1}) on the $C^*$ algebra $\aa_\r$ generated by the creation and annihilation operators $\{a(g), a^*(g)\ :\ g\in L^2(\rx^3,\d^3k)\}$ (called the Canonical Anticommutation Relation (CAR) algebra). We refer the reader to Appendix \ref{appformfactors} for more information on this representation of the CAR algebra.

The Hilbert space \fer{a1} supports in particular all local modifications $\omega$ of $\omega_{\rm ref}$. Such an $\omega$ has the form 
\begin{equation}
\omega(A) = \scalprod{\psi_\omega}{\pi(A)\psi_\omega},
\label{a3.1}
\end{equation}
for all $A\in\aa=\aa_\s\otimes\aa_\r$, for some $\psi_\omega\in\h$, where $\pi:\mm\rightarrow {\cal B}(\h)$ is the representation map. 

The dynamics of the system generated by $H^j$, \fer{i3}, is represented on $\h$ by a {\it Liouville operator} $L^j$:
\begin{equation}
\pi\big(\e^{\i tH^j} A\e^{-\i t H^j}\big) = \e^{\i tL^j}\pi(A)\e^{-\i tL^j},
\label{a4}
\end{equation}
for all $A\in\aa$ and all $t\in\rx$. Consequently, \fer{i2} is represented on $\h$ as
\begin{equation}
\e^{\i t_1 L^1}\cdots \e^{\i t_N L^N}\pi(A)\e^{-\i t_NL^N}\cdots \e^{-\i t_1 L^1}.
\label{a4.1}
\end{equation}(Of course, we understand that the thermodynamic limit has been performed.) The Liouville operators have the form (see Appendix \ref{appformfactors})
\begin{equation}
L^j=L_0^j + \lambda_jV^j,
\label{8}
\end{equation}
where 
\begin{equation}
L_0^j= L_\s^j +L_\r.
\label{10.1}
\end{equation}
Here, 
\begin{equation}
L^j_\s=H^j_\s\otimes\bbbone -
\bbbone\otimes H^j_\s
\label{11}
\end{equation}
acts on $\h_\s\otimes\h_\s$ and 
\begin{equation}
L_\r=\d\Gamma(u) 
\label{9}
\end{equation}
is the second quantization of multiplication by $u\in\rx$ acting on ${\cal F}$. 
The interaction operator 
\begin{equation}
V^j = \pi(v^j)
\label{a3}
\end{equation}
belongs to the $C^*$ algebra $\pi(\aa)$.

It is useful and standard to consider the weak closure of $\aa$,
\begin{equation}
\mm=\big(\aa_\s\otimes\aa_\r\big)'' = {\cal B}(\h_\s)\otimes\bbbone_\s\otimes{\frak A}_\r''.
\label{a4.2}
\end{equation} 
$\mm$ is a von Neumann algebra acting on $\h$. We introduce the reference state
\begin{equation}
\psiref = \psi_\s\otimes\psi_\r,
\label{m3'}
\end{equation}
where $\psi_\s\in\h_\s\otimes\h_\s$ represents the trace state of $\s$, and where $\psi_\r\in\cal F$ is the vacuum vector of $\cal F$, representing the equilibrium state of $\r$. The vector $\psiref$ is cyclic and separating for the von Neumann algebra $\mm$, and we denote by $J=J_\s\otimes J_\r$, $\Delta=\bbbone_\s\otimes\bbbone_\s\otimes\Delta_\r$ the modular conjugation and the modular operator of the pair $(\mm,\psiref)$ (see also \cite{BR,JP3}). It follows from the form of the interaction, \fer{i5} and Assumption (R) in Section \ref{introsect} that $\Delta^{1/2} V^j\Delta^{-1/2}\in\mm$ for all $j$.

\subsection{Proof of \fer{ii1}}
\label{ssii1}

Our main result on the piecewise constant dynamics is the following. 
\begin{theorem}[Dominant paths]
\label{mmthm1} There is a constant $c>0$ s.t. if $\max_j|\lambda_j|<c$, then we have the following. Let $A\in\mm_\s$ be any observable of $\s$, and let $\psi_0$ be any initial state of $\s+\r$, given by $B\psiref$ some $B\in\mm_\s'$. Then 
\begin{eqnarray}
\lefteqn{
\scalprod{\psi_0}{\e^{\i t_1L^1}\cdots \e^{\i t_N L^N} A\e^{-\i t_N L^N}\cdots \e^{-\i t_1L^1}\psi_0 }}\label{mm23} \\
&& =\sum_{r_1,\ldots,r_N} \e^{\i \sum_{j=1}^N t_j\varepsilon^j(r_j)}\scalprod{\psi_0}{B\,\Pi_0 (r_1,\ldots,r_N) A\psiref}+O(\max_j|\lambda_j|),
\nonumber
\end{eqnarray}
where the $\varepsilon^j(r)$ are the resonance eigenvalues (see \fer{i8} and also Propositon \ref{mprop3}). 
The error term in \fer{mm23} is uniform in the $t_j\geq 0$. Let $\eta^j(r)$, $\widetilde\eta^j(r)$ be the resonance eigenvectors (see \fer{i11} and also \fer{mm22}). Then
\begin{eqnarray}
\lefteqn{
\Pi_0(r_1,\ldots,r_N)}\nonumber\\
&& = |\eta^1(r_1)\rangle\scalprod{\widetilde\eta^1(r_1)}{\eta^2(r_2)}\cdots\scalprod{\widetilde\eta^{N-1}(r_{N-1})}{\eta^N(r_N)}\langle\widetilde\eta^N(r_N)|.
\label{mm24}
\end{eqnarray}
\end{theorem}

{\it Remarks.\ } 1. We think that a more detailed analysis of the remainder term in \fer{mm23} would yield an estimate $O(\max_j|\lambda_j|)$ uniformly in $N$, but we do not prove this here.

2. Theorem \ref{mmthm1} implies formula \fer{ii1}. 

\medskip
The remaining part of this section is devoted to the proof of Theorem \ref{mmthm1}. We build up the proof in several steps.

\subsubsection{Passage from the operators $L^j$ to the operators $K^j$}

Let $(V')^j$ be any operator belonging to the commutant $\mm'$,\footnote{The commutant of $\mm$, \fer{a4}, is defined as $\mm'=\{ A\in{\cal B}(\h)\ :\ AB=BA \ \forall B\in\mm\}$.} and set $K^j=L^j+\lambda_j (V')^j$ (with domain
$\dom(K^j)=\dom(L^j)=\dom(L_0^j)$). We define the
operator $\e^{\i tK^j}$, $t\in\rx$, via the operator-norm convergent
Dyson series
\begin{equation}
\e^{\i tK^j} = \sum_{k\geq 0} \lambda_j^k \int_0^t \d
s_1\cdots\int_0^{s_{k-1}} \d s_k (V')^j(s_k)\cdots (V')^j(s_1)\e^{\i t L^j},
 \label{mb1}
\end{equation}
where $(V')^j(s)=\e^{\i s L^j}(V')^j\e^{-\i sL^j}$. 

Since $J\Delta^{1/2}A\psiref =A^*\psiref$ for all $A\in\mm$, and since $V^j=(V^j)^*$, we see that upon choosing
$$
(V')^j=-J\Delta^{1/2} V^j J\Delta^{1/2},
$$
we have $K^j\psiref=0$ and hence 
$$
\e^{\i tK^j}\psiref=\psiref.
$$
Note that $J\Delta^{1/2} =\Delta^{-1/2}J$, and that $J\mm J=\mm'$, so that $(V')^j\in\mm'$ indeed.  

\begin{proposition}
\label{propkl}
We have 
\begin{equation}
\e^{\i tK^j}A\e^{-\i tK^j} = \e^{\i tL^j} A\e^{-\i tL^j},
 \label{mb2}
\end{equation}
for all $t\in\rx$, $A\in\frak M$.
\end{proposition}

{\it Proof.\ } It is easy to verify directly that for $\phi\in\dom(L_0^j)$, we have $\frac{\d}{\d t}
\e^{\i tK^j}\phi = \e^{\i tK^j}K^j\phi$. We write \fer{mb1} as $\e^{\i
tK^j}= S'(t)\e^{\i tL^j} =\e^{\i tL^j} R'(t)$, where $S'(t)$ is given by the series on the r.h.s. of \fer{mb1}, and
$R'(t)=\e^{-\i tL^j} S(t)\e^{\i tL^j}$. Both $S'(t)$ and $R'(t)$ belong to the commutant ${\frak M}'$. Consequently, we have for all $A\in\frak M$
$$
\e^{\i tK^j} A\e^{-\i tK^j} = S'(t)\e^{\i tL^j} A\e^{-\i tL^j} R'(-t)= \e^{\i tL^j} A\e^{-\i tL^j}\ S'(t) R'(-t).
$$
Furthermore, $S'(t)R'(-t) = \e^{\i tK^j}\e^{-\i tL^j}\e^{\i tL^j}\e^{-\i tK^j}=\bbbone$, and thus we obtain \fer{mb2}.\hfill $\square$

\subsubsection{Resolvent representation of propagators}

\begin{proposition}
\label{propresrep}
Let $A\in\mm$ and $\psi\in\h$. We have for $t\geq 0$
\begin{equation}
\scalprod{\psi}{\e^{\i tK^j} A\psiref} = \frac{-1}{2\pi\i} \int_{{\rx}-\i\gamma}  \e^{\i zt}
\scalprod{\psi}{(K^j-z)^{-1} A\psiref} \d z,
\label{a6}
\end{equation}
if $\gamma> C|\lambda_j|$ for some $C>0$.
\end{proposition}

{\it Proof.\ }
The function $t\mapsto \scalprod{\psi}{\e^{\i tL^j} A\e^{-\i tL^j}\psiref} = \scalprod{\psi}{\e^{\i tK^j} A\psiref}$ is
bounded and continuous in $t\in \rx$. It follows that
\begin{eqnarray}
\scalprod{\psi}{\e^{\i tK^j} A\psiref} &=& \frac{1}{2\pi\i}
\int_{\gamma -\i\infty}^{\gamma+\i\infty} \left[
\e^{zt}\int_0^\infty \e^{-z s} \scalprod{\psi}{\e^{\i sK^j}
A\psiref} \d s\right] \d z\nonumber\\
&=&\frac{1}{2\pi\i} \int_{\gamma -\i\infty}^{\gamma+\i\infty}
\e^{zt} \scalprod{\psi}{\i (K^j+\i
z)^{-1} A\psiref} \d z\nonumber\\
&=&\frac{-1}{2\pi\i} \int_{{\rx}-\i\gamma}  \e^{\i zt}
\scalprod{\psi}{(K^j-z)^{-1} A\psiref} \d z.
 \label{mb3}
\end{eqnarray}
In the first step, we use the Laplace inversion
theorem (see e.g. \cite{W}, Chapter II, Theorem 9.2) and in the second step we integrate the propagator to
obtain the resolvent. \hfill $\square$

\subsubsection{Analytic continuation of matrix elements}

For $\theta\in\rx$ we define the unitary group $U_\theta=\e^{\theta \d\Gamma(\partial_u)}$ on ${\cal F}$, (translation in the radial variable $u$, see \fer{10}), and we set
\begin{equation}
L^j_0(\theta) = U_\theta L_0^j U_\theta^{-1}
\label{m1}
\end{equation}
An easy calculation gives $L^j_0(\theta)= L_\s^j+L_\r +\theta N$, where $N=\d\Gamma(\bbbone)$ is the number operator on $\cal F$. Accordingly, we define in the same way
\begin{eqnarray}
K^j(\theta) &=& L^j(\theta)+\lambda_j I^j(\theta),\\
I^j(\theta) &=& V^j(\theta) +(V')^j(\theta).
\label{m1.1}
\end{eqnarray}

Let $\theta_0>0$ be fixed and define the strip
\begin{equation}
{\cal S}_{\theta_0}=\{\theta\in\cx\ :\ |\Im\theta|<\theta_0\}.
\label{m1.2}
\end{equation}
In accordance with analytic spectral deformation theory, we assume the following analyticity condition.
\begin{itemize}
\item[(C1)] $\theta\mapsto I^j(\theta)$ has an analytic continuation
as a map from ${\cal S}_{\theta_0}$ to the bounded operators on
$\h$, and $\sup_{\theta\in{\cal S}_{\theta_0}}\|
I^j(\theta)\|=C<\infty$.
\end{itemize}

\begin{proposition}
\label{analyticityformfactors}
If the form factors $g^j$ satisfy Assumption (R) in Section \ref{introsect}, then Condition (C1) is satisfied.
\end{proposition}

The proof of this proposition is not hard (one examines directly the explicit expression for $I^j(\theta)$, see also Appendix \ref{appformfactors}). The following is the key technical result of the analytic deformation method.

\begin{proposition}
\label{mprop2} Take $z$ with $\Im z<-C|\lambda_j|$, where $C$ is the constant in Condition (C1) above. The map $\theta\mapsto
(K^j(\theta)-z)^{-1}$ has an extension from $\theta\in\rx$ to $0\leq \Im\theta <
\theta_0$. This extension (denoted by the same symbol) is an
analytic map from $\{\theta\in\cx: 0<\Im\theta<\theta_0\}$ to the
bounded operators of $\h$, and it is continuous in the operator
norm as $\Im\theta\downarrow 0$, at all $\theta\neq 0$.
\end{proposition}

{\it Proof of Proposition \ref{mprop2}.\ } We fix the index $j$ and omit it from the notation. Let $\theta\in\rx$. On
$\dom(L_0)\cap\dom(N)$ we have $L_0(\theta) = L_0+\theta N$ and so, by (C1), $K(\theta) = L_0 +\theta N +\lambda I(\theta)$ 
has analytic continuation to $\theta\in{\cal S}_{\theta_0}$ (as a closed operator on $\dom(L_0)\cap \dom(N)$). The spectrum of the normal
operator $L_0+\theta N$ consists of real eigenvalues $e\in{\rm
spec}(L_\s)$ and of horizontal lines $\{n\theta+\rx\ :\ n\in\mathbb N\}$ of continuous
spectrum. (Note that $L_0+\theta N$ is the sum of two commuting self-adjoint
operators.) For $\Im \theta\neq 0$ the eigenvalues $e$ are
isolated. It follows from standard perturbation theory that the
spectrum of $K(\theta)$ lies within a distance of
$|\lambda|\,\|I(\theta)\|$ from that of $L_0+\theta
N$.

{} For $0\leq \Im\theta<\theta_0$, all $\Im z<-C|\lambda|$ (with $C$ as in condition (C1)) belong to the resolvent set of $L_0+\theta N$, as well as to the resolvent set of $K(\theta)$. For such $z$ we express the resolvent using the
norm-convergent Neumann series
$$
(K(\theta)-z)^{-1} = (L_0+\theta N-z)^{-1}\sum_{n\geq 0}(-\lambda)^n [I(\theta)\,(L_0+\theta N-z)^{-1}]^n.
$$
It follows that for all $z$ with $\Im z<-C|\lambda|$,
\begin{enumerate}
\item[1*.]  $N(K(\theta)-z)^{-1}$ is bounded for all nonzero $\theta$
with $0\leq \Im\theta <\theta_0$,

\item[2*.] $(K(\theta)-z)^{-1}$ is bounded uniformly in $\theta$ s.t.
$0\leq \Im\theta<\theta_0$,

\item[3*.] ${\rm Ran}(K(\theta)-z)^{-1}\subset \dom(L_0)\cap \dom(N)$
for all nonzero $\theta$ s.t. $0\leq \Im\theta<\theta_0$.
 \end{enumerate}
Let $\theta$ be s.t. $0<\Im\theta<\theta_0$ and take $\Delta
\theta$ to be small so that $0<\Im(\theta+\Delta\theta)<\theta_0$.
The resolvent identity gives
\begin{eqnarray}
\lefteqn{ (K(\theta
+\Delta\theta)-z)^{-1} - (K(\theta)-z)^{-1}
=}\nonumber\\
&&(K(\theta +\Delta\theta)-z)^{-1} \big[ \Delta\theta \, N +\lambda I(\theta+\Delta\theta) -\lambda
I(\theta) \big]
(K(\theta)-z)^{-1}.
 \label{mb7}
\end{eqnarray}
 Relation \fer{mb7} (together with the above points 1* and 2*.) shows that $\theta\mapsto
(K(\theta)-z)^{-1}$ is continuous on $\{\theta\in\cx\ :\ 0<\Im\theta<\theta_0\}$ in the topology of bounded
operators. Furthermore,
\begin{eqnarray}
\lefteqn{
 \frac{(K(\theta +\Delta\theta)-z)^{-1} -
(K(\theta)-z)^{-1}}{\Delta\theta}}\nonumber\\
&& = (K(\theta +\Delta\theta)-z)^{-1}\big[ N +\lambda
X(\theta,\Delta\theta)\big] (K(\theta)-z)^{-1},
 \label{mb8}
\end{eqnarray}
where $\lim_{\Delta\theta\rightarrow 0} X(\theta,\Delta\theta) =
\partial_\theta I(\theta)$. Combining \fer{mb7}, \fer{mb8} and points 1*, 2* above, we
see that for all $\Im z<-C|\lambda|$ and $\theta$ s.t. $0<\Im
\theta<\theta_0$,
$$
\partial_\theta (K(\theta)-z)^{-1} = (K(\theta)-z)^{-1}\big[ N
+\lambda \partial_\theta I(\theta) \big] (K(\theta)-z)^{-1},
$$
the r.h.s. being a bounded operator.

We now show that $(K(\theta)-z)^{-1}$ is continuous as $\theta
=x+\i y\rightarrow x\in\rx\backslash\{0\}$, $y>0$. The resolvent
identity gives
 \begin{eqnarray}
\lefteqn{(K(x+\i y)-z)^{-1} = }\nonumber\\
&&(K(x)-z)^{-1} +(K(x+\i y)-z)^{-1} \big[ K(x)-K(x+\i y)\big]
(K(x)-z)^{-1}.
 \label{mb9}
\end{eqnarray}
We rewrite \fer{mb9} in the form
$$
(K(x+\i y)-z)^{-1}[\bbbone -W(y)] = (K(x)-z)^{-1},
$$
where 
$$
W(y):= [K(x)-K(x+\i y)](K(x)-z)^{-1}\longrightarrow 0
$$
in operator norm, as $y\rightarrow 0$ (here it is important that
$x\neq 0$). It follows that
$$
\lim_{y\rightarrow 0_+} (K(x+\i y)-z)^{-1} = \lim_{y\rightarrow 0_+}
(K(x)-z)^{-1}[\bbbone-W(y)]^{-1} = (K(x)-z)^{-1}.
$$
This completes the proof of Proposition \ref{mprop2}.\hfill
$\square$

\subsubsection{Separating pole contributions}
\label{sectpolecontr}

{}For $A\in\mm_\s$ and $0\leq j\leq N-1$ we set 
\begin{equation}
A_j = \e^{\i t_{j+1}L^{j+1}}\cdots \e^{\i t_N L^{N}} A\e^{-\i t_N L^{N}} \e^{-\i t_{j+1}L^{j+1}},
\label{mm15}
\end{equation}
and we define $A_N=A$. Let $B$ be the operator in the commutant $\mm_\s'$ satisfying $\psi_0=B\psiref$, and denote by $P_{\psi_\r}=|\psi_\r\rangle\langle\psi_\r|$ the orthogonal projection onto $\h_\s\otimes\h_\s\otimes \cx\psi_\r$. If all $\lambda_j=0$, then the dynamics of $\s$ and $\r$ decouple, and we have $A_j\in\mm_\s\otimes\bbbone_\r$, and thus $A_j\psiref\in\ran P_{\psi_\r}$. The following result follows from an easy perturbation expansion.
\begin{lemma}
\label{mmlemma1}
Set $\overline{P}_{\!\!\psi_\r}=\bbbone-P_{\psi_\r}$. We have 
$$
\left\|\overline{P}_{\!\!\psi_\r} A_j \psiref\right\|\leq C_j \max_j|\lambda_j|\ ||A||.
$$ 
\end{lemma}
Using this result, we arrive at the representation
\begin{equation}
\scalprod{\psi_0}{\e^{\i t_1L^1} A_1 \e^{-\i t_1L^1}\psi_0 }= \scalprod{\psi_0}{B\e^{\i t_1K^1} A_1 \psiref } =  \scalprod{\psi_0}{B\e^{\i t_1K^1} P_{\psi_\r} A_1 \psiref } +R_1,
\label{mm16}
\end{equation}
where $\|R_1\|\leq C \max_j |\lambda_j|$ (with a constant $C$ depending on $N$ and $||A||$). The scalar product term on the right side of \fer{mm16} can now be treated as
\begin{eqnarray*}
\scalprod{\psi_0}{B\e^{\i t_1K^1} P_{\psi_\r} A_1 \psiref } &=& \scalprod{\psi_0}{B\e^{\i t_1K^1} P_{\psi_\r} \e^{\i t_2 K^2} A_2 \psiref }\\
&=& \scalprod{\psi_0}{B\e^{\i t_1K^1} P_{\psi_\r} \e^{\i t_2 K^2} P_{\psi_\r} A_2 \psiref } +R_2,
\end{eqnarray*}
with $||R_2||\leq C\max_j |\lambda_j|$. We iterate this procedure and arrive at
\begin{eqnarray}
\lefteqn{
\scalprod{\psi_0}{\e^{\i t_1L^1}\cdots \e^{\i t_N L^N} A\e^{-\i t_N L^N}\cdots \e^{-\i t_1L^1}\psi_0 }}\nonumber\\
&& = \scalprod{\psi_0}{B  P_{\psi_\r}\e^{\i t_1K^1} P_{\psi_\r}\cdots  P_{\psi_\r}\e^{\i t_N K^N} P_{\psi_\r} A\psiref}+R,
\label{mm1}
\end{eqnarray}
where the remainder term 
\begin{equation}
R = \sum_{j=1}^N R_j
\label{mm19}
\end{equation}
satisfies $||R||\leq C \max_j |\lambda_j|$, with $C$ depending on $N$.

Using the resolvent representation \fer{mb3}, we obtain 
\begin{equation}
P_{\psi_\r}\e^{\i t_j K^j} P_{\psi_\r} = \frac{-1}{2\pi \i}\int_{\rx-\i\gamma} \e^{\i t_j z}  P_{\psi_\r} (K^j-z)^{-1}  P_{\psi_\r} \ \d z.
\label{mm20}
\end{equation}
We now perform spectral deformation in the integrand on the right side of \fer{mm20}. For
$\theta\in\rx$ we have $P_{\psi_\r}U_\theta^* = P_{\psi_\r}= U_\theta P_{\psi_\r}$ and it follows that 
\begin{equation}
P_{\psi_\r} (K^j-z)^{-1} P_{\psi_\r} = P_{\psi_\r} (K^j(\theta)-z)^{-1} P_{\psi_\r}.
 \label{mb4}
\end{equation}
Proposition \ref{mprop2} shows that \fer{mb4} has an extension to values of $\theta$ in $0\leq \Im\theta<\theta_0$, and that this extension is analytic in the open strip $0 < \Im\theta<\theta_0$, and continuous on $\rx\backslash\{0\}$, provided ${\rm Im} z <-C\max_j|\lambda_j|$. However, since \fer{mb4} is constant
for $\theta$ on the real axis, it must actually be constant on the entire region $\{0\leq\Im\theta<\theta_0\}$.\footnote{Apply the Schwarz reflection principle to the analytic function $F(\theta)-F(0)$, where $F(\theta)$ equals \fer{mb4}.} This shows that \fer{mb4} holds for $\Im z<-C\max_j|\lambda_j|$ and $0\leq\Im\theta<\theta_0$. We thus have
\begin{equation}
P_{\psi_\r}\e^{\i t_j K^j} P_{\psi_\r} = \frac{-1}{2\pi \i}\int_{\rx-\i\gamma} \e^{\i t_j z}  P_{\psi_\r} (K^j(\theta)-z)^{-1}  P_{\psi_\r} \, \d z,
\label{mm21}
\end{equation}
for all $\theta$ with $0\leq {\rm Im} \theta<\theta_0$, and where $\gamma > C\max_j|\lambda_j|$. 
We analyze the integral on the r.h.s. of \fer{mm21} in more detail. The following is a standard result \cite{MSB1,MSB2,JP2}.

\begin{proposition}
\label{mprop3}
 Fix $\theta'$ with $0<\theta'<\theta_0$.
There is a constant $c_0>0$ s.t. if $|\lambda|\leq c_0/\beta$ then
the spectrum of $K^j(\theta)$, lying in the complex half-plane $\{z\in\cx:\
\Im z<\theta'/2\}$, is independent of $\theta$ in the region
$\theta'<\Im\theta<\theta_0$. It consists of the distinct isolated
eigenvalues
$$
\big\{\varepsilon^j(e,s):\ e\in{\rm spec}(L^j_\s),
s=1,\ldots,\nu(e)\big\},
$$
where $1\leq \nu(e)\leq{\rm mult}(e)$ counts the splitting of the
eigenvalue $e$ into distinct resonances. Moreover, we have
$\varepsilon^j(e,s)\rightarrow e$ as $\lambda\rightarrow 0$, for all
$s$, and $\Im\varepsilon(e,s)\geq 0$. The continuous spectrum of
$K^j(\theta)$ lies in the region $\{\Im z > 3\theta'/4\}$.
\end{proposition}

We now ``shift'' the path of integration $\rx-\i\gamma$ of the integral in \fer{mm21} to the path $\rx+3\i\theta'/4$ 
in the upper half plane. Thereby we pick 
up contributions (residues) coming from the poles of the integrand. Let $\c^j(e,s)$ be
a small circle around $\varepsilon^j(e,s)$, not enclosing or
touching any other spectrum of $K^j(\theta)$. Define the
generally non-orthogonal Riesz spectral projections
\begin{equation}
Q^j(e,s) = \frac{-1}{2\pi\i} \int_{\c^j(e,s)} (K^j(\theta)-z)^{-1}\d z,
 \label{mb11}
\end{equation}
and the operator
\begin{equation}
Q^j(\infty) = Q^j(\infty,t_j,\theta) = \frac{-1}{2\pi \i} \int_{\rx+3\i\theta'/4}\e^{\i t_jz_j} (K^j(\theta)-z_j)^{-1} \d z_j.
\label{mm3}
\end{equation}
For any vectors $\psi,\phi\in\h$ we have by standard contour deformation of complex integrals
\begin{eqnarray}
\lefteqn{\frac{-1}{2\pi \i} \int_{\rx-\i\gamma}\e^{\i t_j z_j} \scalprod{\psi}{(K^j(\theta)-z_j)^{-1}\phi}\d z_j } \label{mb12}\\
&&= \sum_{e\in{\rm spec}(L^j_\s)}
\sum_{s=1}^{\nu(e)} \e^{\i t_j\varepsilon^j(e,s)}
\scalprod{\psi}{Q^j(e,s)\phi} +\scalprod{\psi}{Q^j(\infty)\phi}.
\nonumber
\end{eqnarray}
The operator $Q^j(\infty)$ reduces to $\e^{\i t_j L_0(\theta)} \overline P_{\!\!\psi_\r}$ for $\lambda_j=0$, and one can show the following result.
\begin{proposition}[\cite{MSB1}]
\label{propmsb1}
We have $\| P_{\psi_\r} Q^j(\infty) P_{\psi_\r}\|\leq C\lambda_j^2 \e^{-3t_j\theta'/4}$.
\end{proposition}
Combining relations \fer{mm1}, \fer{mm21}, \fer{mb12} and Proposition \ref{propmsb1}, we obtain 
\begin{eqnarray}
\lefteqn{
\scalprod{\psi_0}{\e^{\i t_1L^1}\cdots \e^{\i t_N L^N} A\e^{-\i t_N L^N}\cdots \e^{-\i t_1L^1}\psi_0 }}\label{mm4} \\
&& =\sum_{r_1,\ldots,r_N} \e^{\i \sum_{j=1}^N t_j\varepsilon^j(r_j)}\scalprod{B^*\psi_0}{\Pi (r_1,\ldots,r_N) A\psiref}+R',
\nonumber
\end{eqnarray}
where the (multi-)indices $r_j$ are summed over $\{(e,s)\,:\, e\in{\rm spec}(L^j_\s), s=1,\ldots,\nu(e)\}$, with $\varepsilon^j(r_j) = \varepsilon^j(e,s)$ if $r_j=(e,s)$ and where we introduce
\begin{equation}
\Pi(r_1,\ldots,r_N) = Q^1(r_1)\cdots Q^N(r_N),
\label{mm5}
\end{equation}
with $Q^j(r_j)$ given by \fer{mb11}. The remainder term $R'$ in \fer{mm4} satisfies $\|R\|\leq C\max_j|\lambda_j|$ (with $C$ depending on $N$).

As explained in the introduction, we assume that
\begin{itemize}
\item[{\bf (S)}]  Each projection $Q^j(e,s)$ has rank one (for $\lambda_j>0$).
\end{itemize}
This assumption means that all resonance energies $\varepsilon^j(e,s)$ are simple, and it is valid in all our applications. One may modify the results of \cite{MSB2}, where degenerate resonance energies are treated for time-independent Hamiltonians, to eliminate Condition (S). 

Having rank one, the projections are given by
\begin{equation}
Q^j(r_j) = |\chi^j(r_j)\rangle\langle\widetilde\chi^j(r_j)|,\quad r_j=(e,s),
\label{mm6}
\end{equation}
where 
\begin{eqnarray}
K^j(\theta) \chi^j(r_j) &=& \varepsilon^j(r_j)\chi^j(r_j),\label{mm7}\\
{}[K^j(\theta)]^* \widetilde\chi^j(r_j) &=& [\varepsilon^j(r_j)]^*\widetilde \chi^j(r_j),\label{mm8}
\end{eqnarray}
and the resonance eigenvectors are normalized as
\begin{equation}
\scalprod{\chi^j(r_j)}{\widetilde \chi^j(r_j)}=1.
\label{mm9}
\end{equation}

Using perturbation theory (e.g. the Feshbach technique, \cite{BFS,MSB1,MSB2}), one sees that the resonance eigenvectors have the expansion
\begin{equation}
\chi^j(r_j) =\eta^j(r_j)\otimes\psi_\r +O(\lambda_j) \quad\mbox{and}\quad
\widetilde\chi^j(r_j) = \widetilde\eta^j(r_j)\otimes\psi_\r +O(\lambda_j),
\label{mm22}
\end{equation}
where $\psi_\r$ is the vacuum vector of $\cal F$, and where the vectors $\eta^j(r_j)$ and $\widetilde\eta^j(r_j)$, for $r_j=(e,s)$, belong to the eigenspace of $L^j_\s$ associated to the eigenvalue $e$. Let $P^j_e$ be the orthogonal spectral projection of $L^j_\s$ associated to the eigenvalue $e$. We define the {\it level shift operator} $\Lambda^j(e)$ by
\begin{equation}
\Lambda^j(e) = P^j_e I^j \pbar^j(e) 
(\Lbar^j_0 -e+\i 0)^{-1}\pbar^j(e) I^j P^j_e,
\label{mm10}
\end{equation}
where $\pbar^j_e=\bbbone-P^j_e$ and where $\Lbar^j_0=\pbar^j_e L^j_0 \pbar^j_e\upharpoonright {\rm Ran} \pbar^j_e$. 
The vectors $\eta^j(e,s)$ and $\widetilde\eta^j(e,s)$ are eigenvectors of $\Lambda^j(e)$ and its adjoint $[\Lambda^j(e)]^*$,
\begin{equation}
\Lambda^j(e) \eta^j(e,s) = \delta^j(e,s)\eta^j(e,s) \quad\mbox{and}\quad [\Lambda^j(e)]^* \widetilde\eta^j(e,s) = \overline{\delta^j(e,s)} \widetilde\eta^j(e,s),
\label{mm11}
\end{equation}
satisfying the normalization relation
\begin{equation}
\scalprod{\eta^j(e,s)}{\widetilde\eta^j(e,s)}=1.
\label{mm12}
\end{equation} 
The resonance energies have the expansion
\begin{equation}
\varepsilon^j(e,s) = e -\lambda_j^2\delta^j(e,s) +O(\lambda_j^4). 
\label{mm13}
\end{equation}
The proof of Theorem \ref{mmthm1} is now complete by combining expansion \fer{mm22} with \fer{mm4} and \fer{mm5}.

\section{Applications: details and proofs}
\label{appliproofs}
The setting of the applications is given in Section \ref{sectappl}.

\subsection{Regime of overlapping resonances}

The explicit (perturbative) form of the resonance data for the system is given in the following theorem. Recall that $\tau^j$ and $\sigma$ are defined in \fer{mmr30}.

\begin{theorem}[Resonances of $K$]
\label{thmmmr1}
There is a constant $C$ s.t. if $|\lambda|+ \max_j|\Delta^j| < C$, then the resonances of $K^j$ in the region $\{z\in\cx\ : \ {\rm Im}z<\theta_0\}$ are given by 
\begin{eqnarray}
\varepsilon^j(1) &=& 0 \label{mmr3}\\
\varepsilon^j(2) &=& 2\i\sigma +O\big(\lambda^2(|\lambda|+\Deltamax)\big)\label{mmr5}\\
\varepsilon^j(3) &=& \i\sigma+\sigma\sqrt{(\tau^j)^2-1}+O\big(\lambda^2(|\lambda|+\Deltamax)\big)\label{mmr4},\\
\varepsilon^j(4) &=& \i\sigma-\sigma\sqrt{(\tau^j)^2-1}+O\big(\lambda^2(|\lambda|+\Deltamax)\big)\label{mmr4.1},
\end{eqnarray}
where the square root always means the principal branch, with branch-cut on the negative real axis (the argument function takes values in $(-\pi,\pi]$). The resonance eigenvectors $\chi^j(r)$ and $\widetilde\chi^j(r)$, $r=1,\ldots,4$, are given by \fer{mm22}, with
\begin{eqnarray}
\eta^j(1)&=&\widetilde\eta^j(1) = \frac{1}{\sqrt 2}[\varphi_{++}+\varphi_{--}]\label{mmr6}\\
\eta^j(2)&=&\widetilde\eta^j(2) =\frac{1}{\sqrt 2}[\varphi_{++}-\varphi_{--}]\label{mmr7}\\
\eta^j(3) &=& \varphi_{+-}+y_+^j\varphi_{-+}, \qquad \widetilde\eta^j(3) = \alpha_+^j ( \varphi_{+-}+\overline{y_+^j}\varphi_{-+}) \label{mmr8}\\
\eta^j(4) &=& \varphi_{+-}+y_-^j \varphi_{-+}, \qquad \widetilde\eta^j(4) = \alpha_-^j ( \varphi_{+-}+\overline{y^j_-}\varphi_{-+}) \label{mmr9},
\end{eqnarray}
where
\begin{equation}
y_\pm^j = -\i\tau^j \pm \i\sqrt{(\tau^j)^2-1},\quad \alpha_\pm^j = [1+(\overline{y^j_\pm})^2]^{-1}.
\label{mmr10}
\end{equation}
\end{theorem}

It follows from Theorem \ref{thmmmr1} that all resonances are non-degenerate provided $(\tau^j)^2\neq 1$ for all $j$, a condition we assume to hold in this section. Hence condition (S) is satisfied (see introduction as well as Section \ref{sectpolecontr}). We define the transition coefficients
\begin{equation}
T^j(r,r') = \scalprod{\widetilde\eta^j(r)}{\eta^{j+1}(r')},\qquad r,r'=1,\ldots,4, \quad j=1,2,\ldots,
\label{mmr20}
\end{equation}
in terms of which the dominant path $\Pi_0(r_1,\ldots,r_N)$, \fer{mm24}, can be written as
\begin{equation}
\Pi_0(r_1,\ldots,r_N) = \left[\prod _{j=1}^{N-1} T^j(r_j,r_{j+1})\right] |\eta^1(r_1)\rangle\langle\widetilde\eta^N(r_N)|.
\label{mmr31}
\end{equation}
Before proving Theorem \ref{thmmmr1}, we mention that expressions \fer{mmr6}-\fer{mmr10} yield the following result.
\begin{proposition}[Transition coefficients]
\label{proptranscoeff}
We have 
\begin{eqnarray*}
T^j(1,1) = T^j(2,2)&=&1,\\
T^j(3,3) &=& 1+\overline{\alpha_+^j}y_+^j [y_+^{j+1}-y_+^j],\\
T^j(4,4) &=& 1+\overline{\alpha_-^j}y_-^j [y_-^{j+1}-y_-^j],\\
T^j(3,4) &=& \overline{\alpha^j_+} y^j_+[y_-^{j+1}-y_-^j], \\ 
T^j(4,3) &=& \overline{\alpha^j_-} y^j_-[y_+^{j+1}-y_+^j].
\end{eqnarray*}
All other transition coefficients vanish. (Note that if $\Delta^{j+1}=\Delta^j$ then $T^j(r,r')=\delta_{r,r'}$ (Kronecker symbol)).

In the regime of {\rm separated resonances}, where $\tau_{\rm min}:=\min_j|\tau^j|\ggeq 1$, we have 
\begin{eqnarray*}
T^j(3,3), \ T^j(4,4)  &=& 1+\frac{\tau^{j+1}-\tau^j}{2\tau^j} \mp \frac{|\tau^{j+1}|-|\tau^j|}{2\tau^j} +O\big(1/\tau_{\rm min}^2\big),\\
T^j(3,4), \ T^j(4,3)  &=& \frac{\tau^{j+1}-\tau^j}{2\tau^j} \pm \frac{|\tau^{j+1}|-|\tau^j|}{2\tau^j} +O\big(1/\tau_{\rm min}^2\big).
\end{eqnarray*}
In the regime of {\rm overlapping resonances}, where $\tau_{\rm max}:=\max_j|\tau^j|\lless 1$, we have 
\begin{eqnarray*}
T^j(3,3), \ T^j(4,4)  &=& 1\pm \i \frac{\tau^{j+1}-\tau^j}{2}+O\big(\tau_{\rm max}^2\big),\\
T^j(3,4), \ T^j(4,3)  &=&  \pm \i \frac{\tau^{j+1}-\tau^j}{2} +O\big(\tau_{\rm max}^2\big).
\end{eqnarray*}
\end{proposition}

{\it Proof of Theorem \ref{thmmmr1}.\ } Throughout the proof, we consider $j$ fixed and do not display it. The unperturbed Liouville operator $L_0=L_\r$ has a four-fold degenerate eigenvalue at the origin, and absolutely continuous spectrum filling the entire real axis. We have $K=L_0+I$, where
\begin{equation}
I = \frac{\Delta}{2}L_\s + \lambda [V - V'],
\label{mmr3'}
\end{equation}
with $L_\s=\sigma_z\otimes\bbbone-\bbbone\otimes \sigma_z$, $V=\sigma_x\otimes\bbbone_\s\otimes\varphi_\beta(g)$ and $V'=\bbbone_\s\otimes\sigma_x\otimes\widetilde\varphi_\beta(g)$. The spectrally deformed operator, for $\theta\in{\cal S}_{\theta_0}$, is given by
\begin{equation}
K(\theta) =L_0 +\theta N +I(\theta),
\label{mmr12}
\end{equation}
where $N=\d\Gamma(\bbbone)$ is the number operator in $\cal F$, and where 
\begin{equation}
I(\theta) =\frac{\Delta}{2}L_\s+\lambda[V(\theta)+V'(\theta)].
\label{mmr12'}
\end{equation}
We consider $\theta=\i\theta'$, with $\theta'>0$. Let $P$ be the projection onto $\h_\s\otimes\h_\s\otimes\cx\psi_\r$, and let $\pbar=\bbbone-P$. We denote by $\overline T$ the restriction of an operator $T$ to the range of $\pbar$. The estimate $\left\| \pbar (\overline L_\r+\i\theta' \overline N-z)^{-1}\right\|=[{\rm dist}(z,\i\theta'{\mathbb N}^*+\rx)]^{-1}$ implies 
\begin{equation}
\left\| \pbar (\overline L_\r+\i\theta' \overline N-z)^{-1}\right\|=\frac{2}{\theta'},\quad \mbox{for ${\rm Im}z<\theta'/2$}.
\label{mmr13}
\end{equation}
The Neumann series 
\begin{eqnarray}
\lefteqn{
\pbar(\overline{K}(\i\theta')-z)^{-1}\pbar}\nonumber\\
&& = \pbar(\overline L_\r+\i\theta' \overline N-z)^{-1}\pbar\sum_{n\geq 0}(-1)^n\left[ \overline{I}(\i\theta')(\overline L_\r+\i\theta \overline N-z)^{-1}\pbar\right]^n
\label{mmr14}
\end{eqnarray}
converges for all ${\rm Im }z<\theta'/2$, provided that $\|\overline{I}(\i\theta')\|\frac{2}{\theta'}<1$. The latter condition is satisfied for $|\Delta|+|\lambda|< C(\theta_0)$, see the assumptions in Theorem \ref{thmmmr1}. We consider spectral points $z$ with ${\rm Im}z<\theta'/2$. The Feshbach map method \cite{MSB1,MSB2,BFS} tells us that such a $z$ belongs to the spectrum of $K(\i\theta')$ if and only if it belongs to the spectrum of the Feshbach map applied to $K(\i\theta')$,
\begin{equation}
F_{P,z}(K(\i\theta')) = P\left[ K(\i\theta') - I(\i\theta')\pbar (\overline K(\i\theta')-z)^{-1}\pbar I(\i\theta')\right]P.
\label{mmr15'}
\end{equation}
Using \fer{mmr12'} and \fer{mmr14} we obtain
\begin{eqnarray}
\lefteqn{ F_{P,z}(K(\i\theta')) = P\frac{\Delta}{2}L_\s}\label{mmr15} \\ 
&&-\lambda^2 P[V(\i\theta')-V'(\i\theta')]\pbar  (\overline L_\r +\i\theta'\overline N -z)^{-1}\pbar [V(\i\theta')-V'(\i\theta')] P \nonumber\\
&& + O\big(\lambda^2(|\Delta|+|\lambda|)\big).
\nonumber
\end{eqnarray}
Furthermore, the estimate $(\overline L_\r+\i\theta'\overline N-z)^{-1}\pbar = (\overline L_\r+\i\theta'\overline N)^{-1}\pbar+O(|z|/(\theta')^2)$ leads to (fixed $\theta'$)
\begin{equation}
F_{P,z}(K(\i\theta')) = \Lambda +O\big(\lambda^2(|\Delta|+|\lambda|+|z|)\big),
\label{mmr16}
\end{equation}
where 
\begin{equation}
\Lambda = \frac{\Delta}{2} L_\s -\lambda^2\Lambda_\r 
\label{mmr16'}
\end{equation}
is the level shift operator. Here, 
\begin{equation}
\Lambda_\r = P[V-V']\pbar(\overline L_\r+\i 0)^{-1}\pbar [V-V']P.
\label{mmr17}
\end{equation}
We understand \fer{mmr16}, \fer{mmr16'} and \fer{mmr17} as operators acting on ${\rm Ran} P=\cx^2\otimes\cx^2$. In \fer{mmr17} we have eliminated the spectral deformation parameter $\i\theta'$ by analyticity in a standard fashion, replacing it by the (operator norm) limit of $(\overline L_\r+\i\varepsilon)^{-1}\pbar$ as $\varepsilon\downarrow 0$. 

Our next task is to calculate the eigenvalues and eigenvectors of the level shift operator $\Lambda$. Using the explicit form of $V$ and $V'$ we obtain the following result by a standard and straightforward calculation (see also \cite{M,BJM1,MSB1,MSB2}, for instance).
\begin{lemma}
We have $\lambda^2\Lambda_\r = -\i\sigma + \i\sigma\ (\sigma_x\otimes\sigma_x)$, where $\sigma=\frac{\pi}{2}\lambda^2\gamma_0$, with $\gamma_0$ given in \fer{mmr5'}, and where $\sigma_x$ is the Pauli matrix (c.f. \fer{mmr2}). 
\end{lemma}
In the ordered orthonormal basis $\{\varphi_{++},\varphi_{+-},\varphi_{-+},\varphi_{--}\}$ of $\cx^2\otimes\cx^2$, the level shift operator \fer{mmr16'} is represented by the matrix
\begin{equation}
\Lambda = \i\sigma +
\left[
\begin{array}{cccc}
0 & 0 & 0 & -\i\sigma \\
0 & \Delta & -\i\sigma & 0\\
0 &-\i\sigma & -\Delta & 0\\
-\i\sigma & 0 & 0 & 0
\end{array}
\right].
\label{mmr18}
\end{equation}
It is now a simple matter to verify that 
$$
{\rm spec}(\Lambda) =\{ 0,2\i\sigma, \i\sigma [1\mp \i\sqrt{(\Delta/\sigma)^2-1}]\},
$$ 
with corresponding eigenvectors $\eta(1),\ldots,\eta(4)$ given by \fer{mmr6}-\fer{mmr9} (with $j$ fixed). 

It follows from \fer{mmr16} and the isospectrality of the Feshbach map (mentioned before \fer{mmr15'}) that the resonance eigenvalues $\varepsilon(r)$ are given by \fer{mmr3}-\fer{mmr4}. 

One equally easily finds the eigenvectors $\widetilde\eta(1),\ldots,\widetilde\eta(4)$ of the adjoint of \fer{mmr18}.

This completes the proof of Theorem \ref{thmmmr1}. \hfill $\square$

\subsubsection{Single sudden level crossing} 
\label{subsubsslc}

The Hamiltonian is given by \fer{mmr41}. The following is a proof of the expression \fer{mmr45.1} for the transition probability $p_{\rm ge}(t)$. 

The probability $p_{\rm ge}(t)$, for $t>t_{\rm c}$, is given by
\begin{equation}
p_{\rm ge}(t) = \scalprod{\psi_0}{\e^{\i t_{\rm c}L^1}\e^{\i (t-t_{\rm c})L^2} (A\otimes\bbbone_\s\otimes\bbbone_\r) \e^{-\i (t-t_{\rm c})L^2}\e^{-\i t_{\rm c}L^1}\psi_0},
\label{mmr46}
\end{equation}
where $\psi_0=\varphi_{--}\otimes\psi_\r$, and where $A=|\varphi_-\rangle\langle \varphi_-|$. (Note that $\varphi_-$ is the {\it excited} state of $H_\s^2$.). We use Theorem \ref{mmthm1} with $B=\sqrt{2}\ \bbbone_\s\otimes|\varphi_-\rangle \langle\varphi_-|$ (so that $B\psi_{\s}=\varphi_{--}$, where $\psi_\s$ is the trace state of $\s$). The sum in \fer{mm23} has only two terms (since we have only one jump in the Hamiltonian), and we obtain
\begin{eqnarray}
\lefteqn{
p_{\rm ge}(t)}\label{mmr47}\\
&& =\sum_{r_1,r_2} \e^{\i t_{\rm c}\varepsilon^1(r_1) + \i (t-t_{\rm c})\varepsilon^2(r_2)} T(r_1,r_2) \scalprod{\varphi_{--}}{\eta^1(r_1)}\scalprod{\widetilde\eta^2(r_2)}{\varphi_{--}} +O(|\lambda| +\Delta_{\rm max}),
\nonumber
\end{eqnarray}
where the remaider here contains an $O(\Delta_{\rm max})$ term since we carry out perturbation theory in the coupling constant $\lambda$ and the energy spacing $\Delta$ simultaneously (overlapping resonance regime). Theorem \ref{thmmmr1} and \ref{proptranscoeff} then imply that 
\begin{equation*}
p_{\rm ge}(t) = \frac{1}{2} +\frac{1}{2}\e^{\i t_{\rm c}\varepsilon^1(2)+\i(t-t_{\rm c})\varepsilon^2(2)}+O(|\lambda| +\Delta_{\rm max}),
\end{equation*}
which shows  \fer{mmr45.1} for $t>t_{\rm c}$. The proof for $0\leq t<t_{\rm c}$ goes along the same lines (and is actually easier, since there is no jump in the Hamiltonian in this case). \hfill $\square$

\subsubsection{Continuum limit and slow variation expansion ($\tau'(t)$ small)}

\label{subsubcontlimit}

We investigate in this section the continuum limit, as explained in Section \ref{sectappl}.

All quantities in Theorem \ref{thmmmr1} and Proposition \ref{proptranscoeff} can be viewed as depending on continuous time $t$ by the substitution $j\mapsto j\frac tN$. For instance $\varepsilon^j(3) = \varepsilon(j\frac tN,3)$, where 
\begin{equation}
\varepsilon(t,3)=i\sigma +\i\sigma \sqrt{1-\tau(t)^2} +O\big(\lambda^2(|\lambda|+\Delta_{\rm max})\big),
\label{mmr57}
\end{equation}
with $\tau^j=\tau(j\frac tN)$ and $y^j_\pm = y_\pm(j\frac tN)$ with 
$$
y_\pm (t)= -\i\tau(t)\mp \sqrt{1-\tau(t)^2}.
$$
Similarly, the continuous time resonance vectors are denoted by $\eta(t,r)$, $\widetilde\eta(t,r)$; for instance,
\begin{equation}
\eta(t,3) = \varphi_{+-} + y_+(t)\varphi_{-+}, \quad \widetilde\eta(t,3) = \alpha_+(t)\big(\varphi_{+-} + \overline{y}_+(t)\varphi_{-+}\big),
\label{mmr60}
\end{equation}
and so on. 

We define for $s,t\geq0$ the quantities $w(s,t,1)=w(s,t,2)=1$, and
\begin{equation}
w(s,t,3) = \sqrt{\frac{1+y_+(t)^2}{1+y_+(s)^2}},\qquad w(s,t,4) = \sqrt{\frac{1+y_-(t)^2}{1+y_-(s)^2}}.
\label{w}
\end{equation}

It is quite clear (see the proof of Theorem \ref{contlimthm} below for details) that the transition coefficients $T^j(r,r')$ associated to a jump ($r\neq r'$) are of the size 
\begin{equation}
\tau'_{\rm max}:=\sup_{t\geq 0} |\tau'(t)|.
\label{mmr50}
\end{equation}
Now the sum over all paths in \fer{mm23} can be written as a sum over all paths with $k$ jumps, $k=0,\ldots,N-1$. This sum becomes an infinite series in the continuous time limit. Each path with $k$ jumps is of the order of $(\tau'_{\rm max})^k$, and so for small $\tau_{\rm max}'$, one can approximate the series by the first few terms. 

\begin{theorem}[Continuous time limit]
\label{contlimthm}
Let $\eta(t,r)$ and $\widetilde \eta(t,r)$ be the resonance vectors at time $t\geq 0$, where $r=1,\ldots,4$ (see \fer{mmr60}). The continuous time  limit of the main term of \fer{mm23}, 
$
\sum_{r_1,\ldots,r_N} \e^{\i \sum_{j=1}^N t_j\varepsilon^j(r_j)}\scalprod{B^*\psi_0}{\Pi_0 (r_1,\ldots,r_N) A\psiref} 
$, is given by \fer{27}.
\end{theorem}

\bigskip
{\it Proof of Theorem \ref{contlimthm}.\ }  By the mean value theorem we have 
$$
y_\pm^{j+1}-y_\pm^j= \frac tN\left[ -\i\tau'(t_1) \pm\tau'(t_2)\frac{\tau(t_2)}{\sqrt{1-\tau(t_2)^2}} \right], 
$$
where $t_1,t_2\in(jt/N,(j+1)t/N)$. Thus the transition coefficients satisfy
\begin{eqnarray}
T^j(r,r) &=& 1 +O\big(\tau'_{\rm max} t/N\big), \quad \forall j,r \label{mmr51}\\
T^j(3,4) &=& \frac t N \frac{y_+(jt/N)}{1+y_+(jt/N)} \big[ y'_+(jt/N) + O(\tau'_{\rm max}t/N)\big],
\label{mmr52}
\end{eqnarray}
and $T^j(4,3)$ is given by the r.h.s. of \fer{mmr52} with $y_+$ replaced by $y_-$. The remainders $O\big(\tau'_{\rm max} t/N\big)$ are uniform in $j$ and $r$. Relations \fer{mmr51}, \fer{mmr52} show that for ``slow variations'' of $\tau(t)$ the transition coefficient associated to a jump $3\leftrightarrow 4$ is small, proportional to $\tau'(t) t/N$, while no-jump transitions have weight one.

We write the sum over all paths in \fer{mm23} as a sum over all paths with exactly $k$ jumps, where $k=0,1,\ldots,N-1$. (A jump happens if one value of $r_j$ changes to a different value of $r_{j+1}$). There are exactly four paths without any jumps, corresponding to $\Pi_0(r,\ldots,r)$, $r=1,2,3,4$. The paths with a single jump are given by $\Pi_0(r,\ldots,r,r',\ldots r')$, where $(r,r')=(3,4)$ or $(r,r')=(4,3)$, and where the jump takes place at location $k=1,\ldots, N-1$. Note that the only jumps allowed are between $r=3$ and $r=4$, since $T^j(r,r')=0$ if $r\neq r'$ and $r,r'\not\in\{3,4\}$ (see Proposition \ref{proptranscoeff}). 

It is thus clear that we have exactly $2{N\choose k}$ paths with $k$ jumps. The factor $2$ takes into account that $r_1$ can take either of the values $3$ or $4$. For $k$ fixed, the summand in \fer{mm23} is bounded by
\begin{equation}
\left|\e^{\i \sum_{j=1}^N t_j\varepsilon^j(r_j)}\scalprod{B^*\psi_0}{\Pi_0 (r_1,\ldots,r_N) A\psiref}\right| \leq C_0 C_1^k C_2^{N-k},
\end{equation}
where
\begin{eqnarray*}
C_0 &=& \|B^*\psi_0\|\, \|A\|,\\
C_1 &=& \max_j\max_{r\neq r'}|T^j(r,r')|=O\big(\tau'_{\rm max} t/N\big),\\
C_2 &=&\max_{j}\max_{r}|T^j(r,r)| = 1+O\big(\tau'_{\rm max} t/N\big).
\end{eqnarray*}
The last estimates on $C_1$ and $C_2$ follow from \fer{mmr52} and \fer{mmr51}, respectively. The sum over all paths ($N$ fixed) in \fer{mm23} has the upper bound
$$
2C_0\sum_{k=0}^{N-1} {N\choose k} C_1^k C_2^{N-k} = 2C_0\big[ (C_1+C_2)^N - C_1^N\big]\leq 2C_0\big(1+ 2C\tau'_{\rm max}t/N\big)^N,
$$
where $C$ is such that $C_1\leq C\tau'_{\rm max}t/N$ and $C_2\leq 1+C\tau'_{\rm max}t/N$. The limit as $N\rightarrow\infty$ of the r.h.s. $2C_0 \e^{2C\tau'_{\rm max}t}$. This implies that we can truncate in a controlled way the series over the number of jumps obtained in the continuous time limit. If we truncate at $k\leq K$, then the remaining tail of the series is estimated from above by 
$$
2C_0  \sum_{k=K+1}^N {N\choose k} C_1^k C_2^{N-k} \leq 2C_0 (C_1/C_2)^{K+1} \e^{2C\tau'_{\rm max}t} = O\Big((\tau'_{\rm max}t)^{K+1}\ \e^{2C\tau'_{\rm max}t}\Big).
$$
For bounded $t$, the tail of the series is thus $O\big( (\tau'_{\rm max})^{K+1}\big)$.

{\it Contribution of paths without jumps.\ } The products of transition coefficients for the constant paths with $r_j=1,2$ are $1$. Also, 
\begin{equation}
T^1(3,3)\cdots T^{N-1}(3,3)
=
\prod_{j=1}^{N-1} \left[ 1+\frac{y_+(jt/N)}{1+y_+(jt/N)^2} \big(y_+((j+1)t/N)-y_+(jt/N)\big)\right]
\label{mmr55}
\end{equation}
and for $r=4$ the product is given by the r.h.s. of \fer{mmr55} with $y_+$ replaced by $y_-$. By taking the logarithm of \fer{mmr55} the product transforms into a Riemann sum, and so one easily obtains
$$
\lim_{N\rightarrow\infty} T^1(3,3)\cdots T^{N-1}(3,3) = \exp\left\{ \int_0^t \frac{y_+(s)y'_+(s)}{1+y_+(s)^2}\d s\right\} = \sqrt{\frac{1+y_+(t)^2}{1+y_+(0)^2}}.
$$
The limit of the products with $r=4$ is given by the latter square root with $y_+$ replaced by $y_-$. Finally, the limits of the exponential factors in \fer{mm23} are 
\begin{equation}
\lim_{N\rightarrow\infty} \e^{\i \sum_{j=1}^N t_j\varepsilon^j(r_j)} = \e^{\i \int_0^t \varepsilon(s,r_j)\d s},
\label{mmr56}
\end{equation}
where $\varepsilon(s,r)$ is defined as in \fer{mmr57}. This gives the first line in \fer{27}.

{\it Contribution of paths with one jump.\ }
Let $j_0\in\{1,\ldots,N-1\}$ be the location of the jump. We have $r_1=\cdots =r_{j_0}=r$ and $r_{j_0+1}=\cdots=r_{N-1}=r'$, where either $(r,r')=(3,4)$ or $(r,r')=(4,3)$. We treat the first the transition $3\rightarrow 4$. The contribution to the sum in \fer{mm23} is given by 
\begin{equation}
\scalprod{B^*\psi_0}{J_1 \, |\eta(0,3)\rangle\langle \widetilde\eta(t,4)| \,  A\psiref},
\label{mmr62}
\end{equation}
where 
\begin{equation}
J_1=\sum_{j_0=1}^{N-1} \e^{\i\frac tN\sum_{j=1}^{j_0} \varepsilon^j(3)+ \i\frac tN\sum_{j=j_0+1}^{N} \varepsilon^j(4)}\left[\prod_{j=1}^{j_0-1}T^j(3,3)\right] T^{j_0}(3,4) \left[ \prod_{j=j_0+1}^{N-1}T^j(4,4)\right].
\label{mmr61}
\end{equation}
The continuous time limit of $J_1$ is
\begin{equation}
\int_0^t \e^{\i\int_0^s\varepsilon(s',3)\d s' +\i\int_s^t \varepsilon(s',4)\d s'}\ w(0,s,3)\frac{y_+(s)y_-'(s)}{1+y_+(s)^2} w(s,t,4),
\label{mmr62.1}
\end{equation}
with $w$ defined in \fer{w}. The corresponding quantity for the transition $4\rightarrow 3$ is obtained from \fer{mmr62.1} by interchanging the indices $3\leftrightarrow 4$, replacing $y_+$ by $y_-$ and $y'_-$ by $y'_+$. This completes the proof of Theorem \ref{contlimthm}. \hfill $\blacksquare$

\subsection{Regime of isolated resonances}

In the regime of isolated resonances we can use Theorem \ref{mmthm1}.
All that remains is to calculate the eigenvalues $\delta^j$ and eigenvectors $\eta^j$ and $\tilde{\eta}^j$
of the level shift operator  \fer{mm10}.

\begin{theorem}[Resonances]
\label{thmssr1}
Suppose all gaps $\Delta^j$ are numbers, well separated from $0$, independent of $\lambda$.
The resonances are 
\begin{align}
\varepsilon^j(1) &= 0\, ,
&\varepsilon^j(2) &= \lambda_j^2 \delta^j(2) + O(\lambda_j^4)\, ,\label{ssi1}\\
\varepsilon^j(3) &= \Delta^j + \lambda_j^2 \delta^j(3) + O(\lambda_j^4)\, ,
&\varepsilon^j(4) &= -\Delta^j + \lambda_j^2 \delta^j(4) + O(\lambda_j^4)\, ,\label{ssi2}
\end{align}
where, 
\begin{align}
\delta^j(2) &= \i \pi \gamma(\Delta^j)\, ,\label{ssi3}\\
\delta^j(3) &= \frac{\i \pi \gamma(\Delta^j)}{2} - \frac{1}{2} \left\langle {\cal P}\left(\frac{1}{r^2-1}\right), \gamma(r\Delta^j)\right\rangle\, ,\label{ssi4}\\
\delta^j(4) &=  \frac{\i\pi \gamma(\Delta^j)}{2} + \frac{1}{2} \left\langle {\cal P}\left(\frac{1}{r^2-1}\right), \gamma(r\Delta^j)\right\rangle\, ,\label{ssi5}\\
\gamma(r) &= \frac{\sqrt{|r|}}{2}\, \int_{S^2} \d\sigma |g(\sqrt{|r|},\sigma)|^2\label{ssi6}\\
\left\langle {\cal P}\left(\frac{1}{r^2-1}\right), \gamma(r\Delta^j)\right\rangle\, 
&= \lim_{\epsilon \to 0} \int_{-\infty}^{\infty} \frac{1-\chi_{(-\epsilon,\epsilon)}(r^2-1)}{r^2 - 1}\, \gamma(r\Delta^j)\, \d r\label{ssi7}
\end{align}
and the eigenvectors are
\begin{align}
\eta^j(1) &=  \varphi_{++} + \varphi_{--}\, , &
\tilde{\eta}^j(1) &= \frac{1}{e^{-\beta \Delta^j} + 1}[e^{-\beta \Delta^j} \varphi_{++} + \varphi_{--}]\, ,\label{ssi8}\\
\eta^j(2) &= \varphi_{++} - \varphi_{--}\, ,&
\tilde{\eta}^j(2) &=\frac{1}{e^{-\beta \Delta^j}+1}[ \varphi_{++} - e^{-\beta \Delta^j} \varphi_{--}]\, ,\label{ssi9}\\
\eta^j(3) &= \tilde{\eta}^j(3) = \varphi_{+-}\, , &
\eta^j(4) &= \tilde{\eta}^j(4) = \varphi_{-+}\, .\label{ssi10'}
\end{align}
The transition coefficients are $T^j(r,r') = \langle \tilde{\eta}^{j}(r), \eta^{j+1}(r')\rangle$, which equal
\begin{gather}
T^j(1,1) = T^j(2,2) = T^j(3,3) = T^j (4,4) = 1,\label{ssi10}\\
T^j(1,2) = \frac{\sinh(\beta[\Delta^{j+1} - \Delta^j]/2)}{2 \cosh(\beta \Delta^j/2) \cosh(\beta \Delta^{j+1}/2)}.\label{ssi11}
\end{gather}
All other transition coefficients vanish. 
\end{theorem}

{\it Remarks.\ } 1. If we consider the asymptotic regime where $\Delta^j \ll 1$, we obtain $\varepsilon^j(1)=0$,
$\varepsilon^j(2)=2\i\sigma$, $\varepsilon^j(3)=\Delta^j + \i\sigma$,
$\varepsilon^j(4)=-\Delta^j+ \i\sigma$.
(To see this, note that $\langle \frac{1}{r^2-1}, \gamma(r\Delta^j)\rangle = O((\Delta^j)^2)$ in this regime.)
This agrees with Theorem \ref{thmmmr1} if one takes $\tau^j \to \infty$, which signifies $\lambda^2 \ll \Delta^j \ll 1$. The biggest change is the change to $\tau^j$, which previously was just $\Delta^j/\sigma^j$. Now we have 
\begin{equation}
\sigma(\Delta^j) \tau(\Delta^j)
= \Delta^j - \lim_{\epsilon \downarrow 0} \int_{0}^{\infty} \frac{1- \chi_{(-\epsilon,\epsilon)}(r^2-1)}{r^2-1} \gamma(r\Delta^j) \d r.
\label{ssr3}
\end{equation}
The principal value integral appearing in \fer{ssr3} vanishes in the limit $\Delta^j \to 0$, and is therefore part of the remainder in the setting of Theorem \ref{thmmmr1}.

2. Clearly the full analysis of the overlapping region is more involved than taking the limit of the answers from 
the non-overlapping region.
For instance, the limit above does not give the correct answer for the eigenvectors, and hence $\mbox{nalso}$ for the transmission coefficients.

3. Once Theorem \ref{thmssr1} is proved, the calculations leading to \fer{ssr4} and \fer{ssr5} for the single sudden crossing
are done exactly in analogy to Section \ref{subsubsslc}, and in fact are easier.

{\it Proof.}
We have to calculate, and diagonalize the level shift operator \fer{mm10}.
For this purpose we will take $j$ to be fixed, in order to prove \fer{ssi1} -- \fer{ssi10'}.
(The calculation of the transition coefficients follows trivially from these, using the definition \fer{mmr20}.)
We will consider the gap $\Delta^j$ to be positive.
Then $I = V - J_{\r} \Delta_{\r}^{1/2} V \Delta_{\r}^{-1/2} J_{\r}$.
Using \fer{aa6}, \fer{aa7} and \fer{aa9}, we see that
\begin{equation}
I =
\sigma_x \otimes \unity_{\s} \otimes \phi(\tau_{\beta} g)
- \unity_{\s} \otimes \sigma_x \otimes \left[a^*(\tau_{\beta} g) (-1)^N + (-1)^N a(e^{-\beta u} \tau_{\beta}g)\right]
\label{ssi13}
\end{equation}
and the level shift operator at energy $e$ is 
\begin{equation}
\Lambda(e)
= P_e I \bar{P}_e (L_{\r} + L_{\s} - e +\i0)^{-1} \bar{P}_e I P_e,
\label{ssi14}
\end{equation}
with $L_{\s} = \frac{\Delta}{2} (\sigma_z\otimes \unity_{\s} - \unity_{\s} \otimes \sigma_z)\otimes \unity_{\r}$ and 
$L_{\r} = \unity_{\s} \otimes \unity_{\s} \otimes \d\Gamma(u)$.
Also $P_0$ is the projection onto the span of $\{\varphi_{++}\otimes \Omega,\varphi_{--}\otimes\Omega\}$.
So, in this basis
\begin{equation}
\Lambda(0)
= \begin{bmatrix} \Lambda_{11}(0) & \Lambda_{12}(0) \\ \Lambda_{21}(0) & \Lambda_{22}(0)
\end{bmatrix}
\label{ssi15}
\end{equation}
where
\begin{align}
2\Lambda_{11}(0) &= \scalprod{\tau_{\beta} g}{(u-\Delta+\i0)^{-1} \tau_{\beta} g} 
+ \scalprod{e^{-\beta u} \tau_{\beta} g}{(u+\Delta+\i0)^{-1} \tau_{\beta} g} \\
2\Lambda_{22}(0) &= \scalprod{\tau_{\beta} g}{(u+\Delta+\i0)^{-1} \tau_{\beta} g} 
+ \scalprod{e^{-\beta u} \tau_{\beta} g}{(u-\Delta+\i0)^{-1} \tau_{\beta} g} \\
-2\Lambda_{21}(0) &= \scalprod{e^{-\beta u} \tau_{\beta} g}{(u-\Delta+\i0)^{-1} \tau_{\beta} g} 
+ \scalprod{\tau_{\beta} g}{(u+\Delta+\i0)^{-1} \tau_{\beta} g} \\
-2\Lambda_{12}(0) &= \scalprod{e^{-\beta u} \tau_{\beta} g}{(u+\Delta+\i0)^{-1} \tau_{\beta} g} 
+ \scalprod{\tau_{\beta} g}{(u-\Delta+\i0)^{-1} \tau_{\beta} g} 
\end{align}
where the inner products are in $L^2(\rx \times S^2, \d u\, \d \sigma)$.
Noting that $e^{-\beta u} |[\tau_{\beta}g](u,\sigma)|^2 =  |[\tau_{\beta} g](-u,\sigma)|^2$, we see that 
\begin{align}
\Lambda_{11}(0) &= -\Lambda_{12} = \i \scalprod{\tau_{\beta} g}{\Im[(u-\Delta+\i0)^{-1}] \tau_{\beta} g},\\
\Lambda_{22}(0) &= -\Lambda_{21} = \i \scalprod{\tau_{\beta} g}{\Im[(u+\Delta+\i0)^{-1}] \tau_{\beta} g}.
\end{align}
But now we use the well-known formula that $\lim_{\epsilon \downarrow 0} \Im[(u-u_0+\i\epsilon)^{-1}] = -\pi \delta(u-u_0)$, as a distribution
to obtain
\begin{equation}
\begin{split}
\Lambda(0)
&= - \i \pi \int_{S^2} \begin{bmatrix} |\tau_{\beta}  g(\Delta,\sigma)|^2 & -|\tau_{\beta} g(\Delta,\sigma)|^2 \\
-|\tau_{\beta} g(-\Delta,\sigma)|^2 & |\tau_{\beta} g(-\Delta,\sigma)|^2 \end{bmatrix}\, \d\sigma\\
&= -\frac{\i \pi \gamma(\Delta)}{e^{-\beta \Delta} + 1} 
\begin{bmatrix} 1 & -1 \\ - e^{-\beta \Delta} & e^{-\beta \Delta} \end{bmatrix}.
\end{split}
\end{equation}
It is easy to see that the eigenvalues are $\delta(1)=0$ and $\delta(2)$ from \fer{ssi3}, 
with eigenvectors $\eta(1)$, $\eta(2)$, $\tilde{\eta}(1)$ and $\tilde{\eta}(2)$
from \fer{ssi8} and \fer{ssi9}.

For $e=\pm \Delta$ there is only one eigenvector, each. Let us consider $e=\Delta$ whose eigenvector
(left and right) is $\varphi_{+-}$.
In this case the action of $\sigma_x \otimes \unity_{\s}$ and $\unity_{\s} \otimes \sigma_x$
both serve to map $\varphi_{+-}$ to eigenvectors of $L_{\s}$ with eigenvalue $0$.
On the other hand, $e=\Delta$ appears in the resolvent.
So
\begin{equation}
\begin{split}
2\Lambda(\Delta) 
&= \scalprod{\tau_{\beta} g}{(u-\Delta+\i0)^{-1} \tau_{\beta} g} + \scalprod{e^{-\beta u} \tau_{\beta} g}{(u-\Delta+\i0)^{-1} \tau_{\beta} g} \\
&= \scalprod{\tau_{\beta} g}{\left[(u-\Delta+\i0)^{-1} + (-u-\Delta+\i0)^{-1}\right] \tau_{\beta} g}.
\end{split}
\end{equation}
(We again used $e^{-\beta u} |[\tau_{\beta}g](u,\sigma)|^2 =  |[\tau_{\beta} g](-u,\sigma)|^2$.)
Now we use the well-known formula that 
$\lim_{\epsilon \downarrow 0} (u-u_0+i\epsilon)^{-1} = -\i\pi \delta(u-u_0) + {\cal P}(\frac{1}{u-u_0})$, where ${\cal P}(\frac{1}{u-u_0})$
is the Cauchy principle value distribution, ${\cal P}(\frac{1}{u-u_0}) = \lim_{\epsilon \to 0} \frac{\chi_{|u-u_0|>\epsilon}}{u-u_0}$,
where the limit is in the space of distributions.
So in this case we do get a ``Lamb shift'' in addition to the purely imaginary resonance
\begin{equation}
2\Lambda(\Delta)
= \scalprod{\tau_{\beta} g}{[-\i\pi[\delta(u-\Delta) + \delta(u+\Delta)] + {\cal P}(u-\Delta) - {\cal P}(u+\Delta)] \tau_{\beta} g}.
\end{equation}
Consideration of this formula leads to \fer{ssi4}, and \fer{ssi5} then follows by symmetry arguments.

We can treat the case of negative gaps $\Delta^j$ by conjugating by $\sigma_x$ to change to $-\Delta^j$.
This does not affect the resonances but it does affect the eigenvectors. 
However, all that happens is that some vectors $\eta^j(r)$ are multiplied by $-1$ and this is always
accompanied by the same change to the corresponding dual eigenvector $\tilde{\eta}^j(r)$.
This type of gauge transformation does not affect any physical quantities.
\hfill $\square$

\appendix

\section{Araki-Wyss representation}

\subsection{The representation}
\label{awapp}

Here we will outline the Araki-Wyss representation \cite{AW} in order to be self-contained.
We use similar notation to \cite{BJM2} (starting on page 24).
The Araki-Wyss representation of the CAR is a representation on the tensor product of two fermionic Fock
spaces ${\mathcal F}(L^2(\R^3,d^3k)) \otimes {\mathcal F}(L^2(\R^3,d^3k))$, such that
the smoothed-out creation operators are represented by the formula
\begin{equation}
\label{AW1}
a_{\beta}^*(g)\, 
=\, a^*\left(\frac{1}{\sqrt{e^{-\beta |k|^2}+1}} g\right) \otimes \unity 
+ (-1)^{N} \otimes a\left(\frac{1}{\sqrt{e^{\beta |k|^2}+1}} \bar{g}\right)\, ,
\end{equation}
where $N$ is the number operator $N=d\Gamma(\unity)$.
The annihilation operators are $a_{\beta}(g) = [a_{\beta}^*(g)]^*$.
Let us write $\mu_{\beta} = (1+e^{-\beta |k|^2})^{-1}$.
In this context, the modular operator and modular conjugation combine to give
\begin{equation}
\label{AW2}
\begin{split}
\tilde{a}_{\beta}(g)
&= J_{\r} \Delta_{\r}^{1/2} a_{\beta}(g) \Delta_{\r}^{-1/2} J_{\r}\\
&= a^*\left(\frac{e^{\beta |k|^2/2}}{\sqrt{e^{\beta |k|^2}+1}} g\right) (-1)^N \otimes (-1)^N
+ \unity \otimes (-1)^N a\left(\frac{e^{\beta |k|^2/2}}{\sqrt{e^{-\beta |k|^2}+1}} \bar{g}\right),
\end{split}
\end{equation}
and
\begin{equation}
\label{AW3}
\begin{split}
\tilde{a}_{\beta}^*(g)
&= J_{\r} \Delta_{\r}^{1/2} a_{\beta}^*(g) \Delta_{\r}^{-1/2} J_{\r}\\
&= (-1)^N a\left(\frac{e^{-\beta |k|^2/2}}{\sqrt{e^{\beta |k|^2}+1}} g\right)  \otimes (-1)^N
+ \unity \otimes a^*\left(\frac{e^{-\beta |k|^2/2}}{\sqrt{e^{-\beta |k|^2}+1}} \bar{g}\right) (-1)^N.
\end{split}
\end{equation}

\subsection{Regularity of form factors}
\label{appformfactors}

Let $g\in L^2(\rx_+\times S^2,|k|^2\d|k|\, \d\sigma)$ be a form factor represented in spherical coordinates ($\d\sigma$ being the uniform measure on $S^2$). We introduce a new radial coordinate $r=|k|^2$ so that the dispersion relation of the Fermions reads $\omega=|k|^2=r$, i.e., $H_\r=\d\Gamma(r)$ on the Fock space ${\cal F}(L^2(\rx_+\times S^2,\frac{\sqrt r}{2}\d r\, \d\sigma))$ (see \fer{i4}). The Araki-Wyss representation Hilbert space associated to the thermal equilibrium is 
\begin{equation}
{\cal F}\big(L^2(\rx_+\times S^2,\textstyle\frac{\sqrt r}{2}\d r\, \d\sigma)\big) \otimes {\cal F}\big(L^2(\rx_+\times S^2,\textstyle\frac{\sqrt r}{2}\d r\, \d\sigma)\big).
\label{aa1}
\end{equation}
For the purpose of spectral deformation, it is advantageous to use the maps
\begin{equation}
a^\#(f)\otimes\bbbone \mapsto a^\#(f\oplus 0)\qquad (-1)^N\otimes a^\#(f)\mapsto a^\#(0\oplus f)
\label{aa2}
\end{equation}
to define an isometric isomorphism between \fer{aa1} and the Hilbert space
\begin{equation}
{\cal F}\Big( L^2(\rx_+\times S^2, \textstyle\frac{\sqrt r}{2}\d r\, \d\sigma) \oplus L^2(\rx_+\times S^2,\textstyle\frac{\sqrt r}{2}\d r\, \d\sigma)\Big).
\label{aa3}
\end{equation}
In \fer{aa2}, $N$ is the number operator. A further isometric isomorphism between \fer{aa3} and 
\begin{equation}
{\cal H}_\r={\cal F}\Big( L^2(\rx\times S^2, \d u\, \d\sigma)\Big)
\label{aa4}
\end{equation}
(see also \fer{10}) is induced by such an isomorphism between the one-particle spaces, given by 
\begin{equation}
f\oplus g \mapsto h,\qquad h(u,\sigma) = \frac{|u|^{1/4}}{\sqrt 2}
\left\{
\begin{array}{ll}
f(u,\sigma) & \mbox{if $u\geq 0$},\\
g(-u,\sigma) & \mbox{if $u<0$},
\end{array}
\right.
\label{aa5}
\end{equation}
where $f,g\in L^2(\rx_+\times S^2, \textstyle\frac{\sqrt r}{2}\d r\, \d\sigma)$ and $h\in L^2(\rx\times S^2,\d u\, \d\sigma)$. 

Under these isomorphisms, the field operator $\phi(g)$, in the Araki-Wyss representation, has the expression 
\begin{equation}
\phi_\beta(g) = \phi(\tau_\beta g),
\label{aa6}
\end{equation}
where $\phi$ on the r.h.s. is the field operator on the fermionic Fock space \fer{aa3}, and where 
\begin{equation}
[\tau_\beta g](u,\sigma) = 
\frac {1}{\sqrt 2}\sqrt\frac{|u|^{1/2}}{e^{-\beta u}+1}
\left\{
\begin{array}{ll}
g(\sqrt u,\sigma), &\mbox{if $u\geq 0$},\\
\overline{g}(\sqrt{-u},\sigma), & \mbox{if $u<0$}.
\end{array}
\right.
\label{aa7}
\end{equation}
Using the explicit transformations introduced above, the expressions \fer{8}-\fer{a3} are easily found. Furthermore, it is not hard to analyze the explicit action of the deformation transformation $U_\theta$ (see before \fer{m1}). For instance, for $h\in L^2(\rx\times S^2,\d u\d\sigma)$ and $\theta\in\rx$, we have
\begin{equation}
U_\theta \phi(h) U_\theta^* = \phi(\e^{\theta\partial_u}h),
\label{aa8}
\end{equation}
where $(\e^{\theta\partial_u}h)(u,\sigma) = h(u+\theta,\sigma)$. One can now combine \fer{aa8} with $h=\tau_\beta g^j$ (see \fer{aa6}, \fer{aa7}) and check that due to Condition (R) in Section \ref{introsect} (with $\alpha(u)=1$), $\theta\mapsto U_\theta \phi(h) U_\theta^*$ admits an analytic continuation into a strip $|\Im\theta|<\theta_0$. One can proceed similarly to check analyticity of $(V^j)(\theta)$ by using the explicit form $\Delta_\r=\e^{-\beta L_\r}$ for the modular operator of $\r$. 
For reference in the body of the paper, let us note explicitly that
\begin{equation}
\label{aa9}
J_{\r} \Delta_{\r}^{1/2} \phi_{\beta}(g) J_{\r} \Delta_{\r}^{1/2}
= a^*(\tau_{\beta} g) (-1)^N + (-1)^N a(e^{-\beta u} \tau_{\beta}g)\, ,
\end{equation}
in this representation ${\cal F}( L^2(\rx\times S^2, \d u\, \d\sigma))$.
This can also be seen by comparing to \fer{AW2} and \fer{AW3}.

\end{document}